\documentclass{nature}
\usepackage{graphicx}
\usepackage{amssymb}
\usepackage{dcolumn}
\usepackage{bm}
\usepackage{epstopdf}
\usepackage{epsfig}
\usepackage{color}
\usepackage[colorlinks=true, letterpaper=true, pdfstartview=FitV, linkcolor=blue, citecolor=blue, urlcolor=blue]{hyperref}
\newcommand{\onlinecite}[1]{\hspace{-1 ex} \nocite{#1}\citenum{#1}}

\begin{document}

\title{Electronic, Dielectric, and Plasmonic Properties of Two-Dimensional Electride Materials X$_2$N (X=Ca, Sr): A First-Principles Study}

\author{Shan Guan$^{1,2}$, Shengyuan A. Yang$^{2,\star}$, Liyan Zhu$^3$, Junping Hu$^1$, Yugui Yao$^{1,\star}$}

\maketitle

\begin{affiliations}
\item School of Physics, Beijing Institute of Technology, Beijing 100081, China
\item Research Laboratory for Quantum Materials and EPD Pillar, Singapore University of Technology and Design, Singapore 487372, Singapore
\item School of Physics and Electronic \& Electrical Engineering, Huaiyin Normal University, Huaian 223300, China

$^\star$e-mail: shengyuan\_yang@sutd.edu.sg; ygyao@bit.edu.cn
\end{affiliations}

\begin{abstract}
Based on first-principles calculations, we systematically study the electronic, dielectric, and plasmonic properties of two-dimensional (2D) electride materials X$_2$N (X=Ca, Sr). We show that both Ca$_2$N and Sr$_2$N are stable down to monolayer thickness. For thicknesses larger than 1-monolayer (1-ML), there are 2D anionic electron layers confined in the regions between the [X$_2$N]$^+$ layers. These electron layers are strongly trapped and have weak coupling between each other. As a result, for the thickness dependence of many properties such as the surface energy, work function, and dielectric function, the most dramatic change occurs when going from 1-ML to 2-ML. For both bulk and few-layer Ca$_2$N and Sr$_2$N, the in-plane and out-of-plane real components of their dielectric functions have different signs in an extended frequency range covering the near infrared, indicating their potential applications as indefinite media. We find that bulk Ca$_2$N and Sr$_2$N could support surface plasmon modes in the near infrared range. Moreover, tightly-bounded plasmon modes could exist in their few-layer structures. These modes have significantly shorter wavelengths (~few tens of nanometers) compared with that of conventional noble metal materials, suggesting their great potential for plasmonic devices with much smaller dimensions.
\end{abstract}

Electrides are a special kind of ionic solids with cavity-trapped electrons serving as the anions\cite{Dye2003,Dye2009}. These electrons are spatially separated from the cations in a regular crystalline array, and are not bound to any particular atom, molecule, or bond. The first crystalline electride, Cs$^+\cdot$(18-crown-6)$_2\cdot$e$^-$, was synthesized in 1983\cite{Ellaboudy1983}, and several other electrides have been successfully discovered or predicted later on\cite{Huang1988,Dye1990,Matsuishi2003,Li2003,Toda2007,Kim2012,Fang2000,Kitano2012,Picard2010,Gatti2010,Miao2014}. The properties of electrides are closely connected to the topology and geometry of the cavities which confine the anionic electrons\cite{Dye2009}. The early examples of electride demonstrate confinement of zero-dimensional cavities or one-dimensional weakly linked channels\cite{Dye2009}. Recently, a new type of electride with two-dimensional (2D) confinement of anionic electrons was discovered in dicalcium nitride (Ca$_2$N) which has a layered structure\cite{Lee2013}.
From formal valence consideration, each unit cell of Ca$_2$N should have one excess electron. Experimental measurements on properties such as the transport parameters and work functions, combined with first-principles calculations, indeed proved that there is a built-in anionic electron layer confined between the calcium layers, which agrees well with the chemical formula [Ca$_2$N]$^+\cdot$e$^-$\cite{Lee2013}. This discovery of 2D electride materials (here 2D refers to the anionic electron confinement topology) has generated great research interest. First-principles calculations have predicted several other 2D electrides\cite{Inoshita2014,Tada2014}, including the other two alkaline earth sub-nitrides Sr$_2$N and Ba$_2$N\cite{Walsh2013}. Recent theoretical work also indicated that the monolayer Ca$_2$N might be mechanically exfoliated from the bulk, meanwhile, its electron confined layer could be maintained, and suitable encapsulation layers were proposed to protect it in ambient environment\cite{Zhao2014}.

Motivated by these recent experimental and theoretical progress on 2D electrides, and also by the surge of research activities on 2D materials in recent years initiated by the discovery of graphene, in this work, we conduct a systematically investigation of the electronic, dielectric, and plasmonic properties of monolayer and few-layer alkaline earth sub-nitrides Ca$_2$N and Sr$_2$N. We find that besides Ca$_2$N, Sr$_2$N is also dynamically stable down to monolayer thickness. Their phonon spectra exhibit characteristic features of 2D materials. For thickness greater than 1-ML, besides the two surface electron bands, there are additional bands crossing the Fermi energy which are due to the anionic electron layers confined in the 2D interlayer regions between the [X$_2$N]$^+$ layers. These interlayer bands are lower in energy than the surface bands and are nearly degenerate, indicating that they are only weakly coupled. The thickness dependence of properties such as surface energy, work function, and dielectric function are analyzed. The change in property with thickness is most dramatic when going from 1-ML to 2-ML, which is associated with the appearance of the first interlayer electron band. We find that Sr$_2$N has a lower surface energy and lower work function compared with Ca$_2$N. Due to their intrinsic structural anisotropy, these layered materials have highly anisotropic dielectric functions. In particular, the in-plane and out-of-plane components of dielectric function can have different signs, which occurs already for the bulk form in the near infrared frequency range. Moreover the dissipation is low in the range for the bulk materials. This shows that bulk Ca$_2$N and Sr$_2$N could be ideal low-loss indefinite media\cite{Smith2003}. We further investigate the possibility of guided surface plasmon modes for these metallic materials and show that bulk Ca$_2$N and Sr$_2$N have good performance comparable to the noble metals but operating at lower frequencies, while their thin film structure can have strongly bounded plasmon modes which have advantage of much shorter wavelength compared to the conventional noble metals. Our findings thus identify a class of promising 2D electride materials and reveal their potentials for future electronics and plasmonics applications.

\section*{Results}
\subsection{Crystal structure, dynamical stability, and surface energy.}
The alkaline earth subnitrides X$_2$N (X=Ca, Sr, and Ba) in their bulk form can be synthesized by a direct solid-state reaction of the elements or through chemical reduction of the corresponding sesquinitrides\cite{Gregory2000}. Their structures consist of alternating ABC-stacked (X-N-X) hexagonal layers, which are in the $R\bar{3}m$ space group (anti-CdCl$_2$-type) with a high $c/a$ ratio, where $a$ and $c$ are the in-plane and out-of-plane unit cell dimensions\cite{Baker2001,Reckeweg2002}. Each (X-N-X) layer is closely packed. For example, the [Ca$_2$N] layer has a thickness of 2.51\textup{\AA} which is much smaller than that of the ordinary fcc Ca (111) layers (3.18\textup{\AA}). This is typically attributed to the ionic bounding in the layer. Meanwhile, the separation between two neighboring layers is relatively large, leading to 2D confined spaces for the anionic electrons\cite{Lee2013}.

\begin{figure}
  \begin{center}
   \epsfig{file=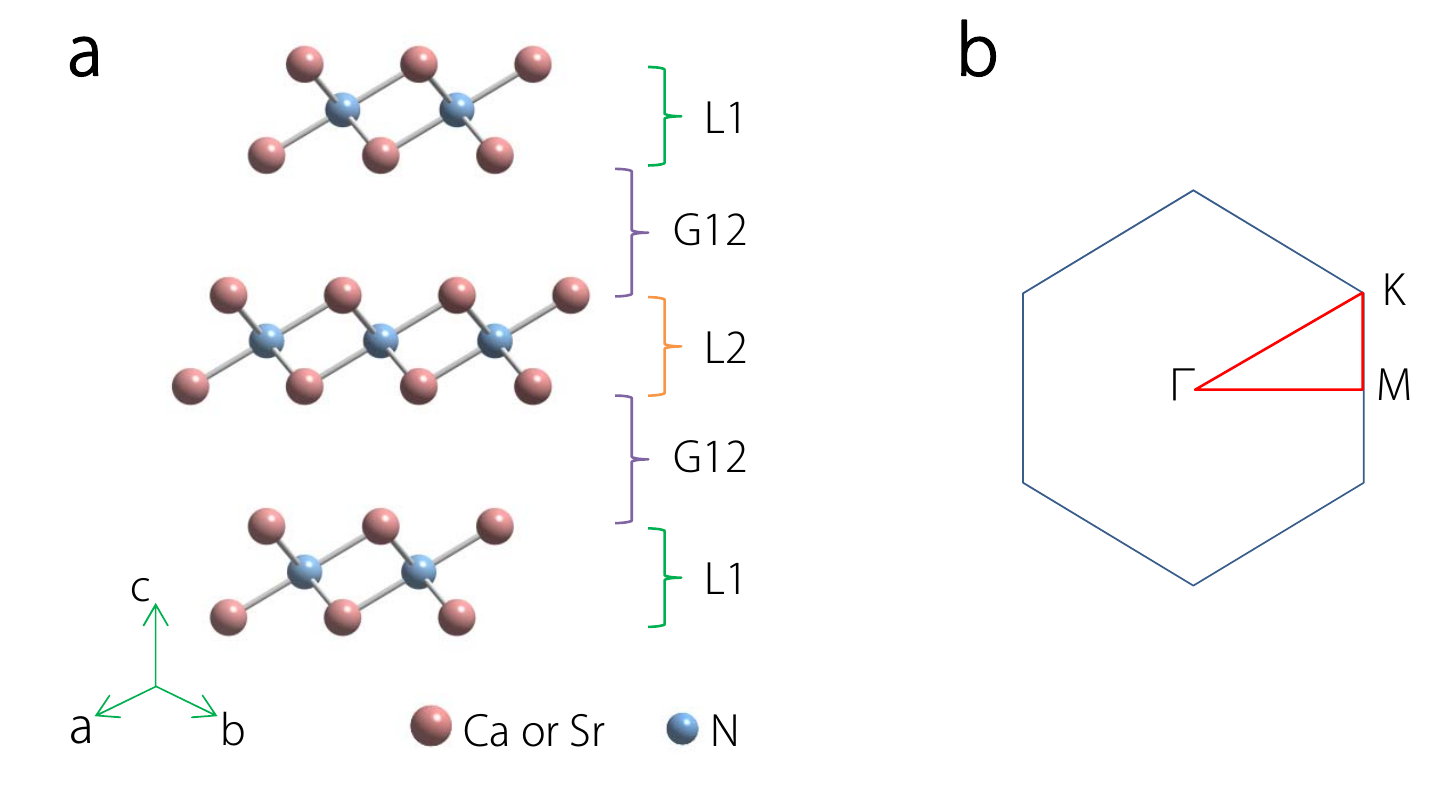,width=14cm}
  \end{center}
  \caption{\label{Fig1} \textbf{a}, Schematic figure of crystal structure of 3-ML X$_2$N (X=Ca, Sr) viewed from [110] direction. It consists of three (X-N-X) unit layers with ABC-stacking. The symbols L1, L2, and G12 refer to the thicknesses of outermost layer, the next outermost layer, and the interlayer gap between L1 and L2, respectively, as explained in the caption of Tabel~\ref{lattice}.
  \textbf{b}, 2D Brillouin zone of X$_2$N (X=Ca, Sr) few-layer structures in which the high symmetry points are labeled.}
\end{figure}

In this work, we investigate the properties of monolayer and few-layer thin films of these subnitrides (here one monolayer (ML) refers to one (X-N-X) unit layer). Since large interlayer separations ($>3.5\textup{\AA}$) usually signal a possible role of long-range van der Waals interaction, the van der Waals corrections was included in the first-principles calculation\cite{Grimme2006}. The calculation details are described in the Method section. As an example, the crystal structure of 3-ML X$_2$N (X=Ca, Sr) is schematically shown in Fig.\ref{Fig1}a. The optimized lattice parameters for isolated Ca$_2$N and Sr$_2$N thin films with thickness from 1-ML to 5-ML are listed in Table~\ref{lattice}. One observes that despite slight variations with respect to the number of layers, the calculated in-plane lattice constants, layer thicknesses, and interlayer distances are close to the experimental values of the corresponding bulk structures (for bulk Ca$_2$N, $a=3.62\textup{\AA}$, layer thickness$=2.51\textup{\AA}$, and layer separation$=3.86\textup{\AA}$; for bulk Sr$_2$N, $a=3.85\textup{\AA}$, layer thickness$=2.71\textup{\AA}$, and layer separation$=4.19\textup{\AA}$)\cite{Reckeweg2002}. Generally, the interlayer spacing is much larger than the layer thickness by about 50\%. Sr$_2$N has a larger interlayer spacing (around 4\textup{\AA}) than Ca$_2$N due to the larger atomic number of Sr.

\begin{figure}
  \begin{center}
   \epsfig{file=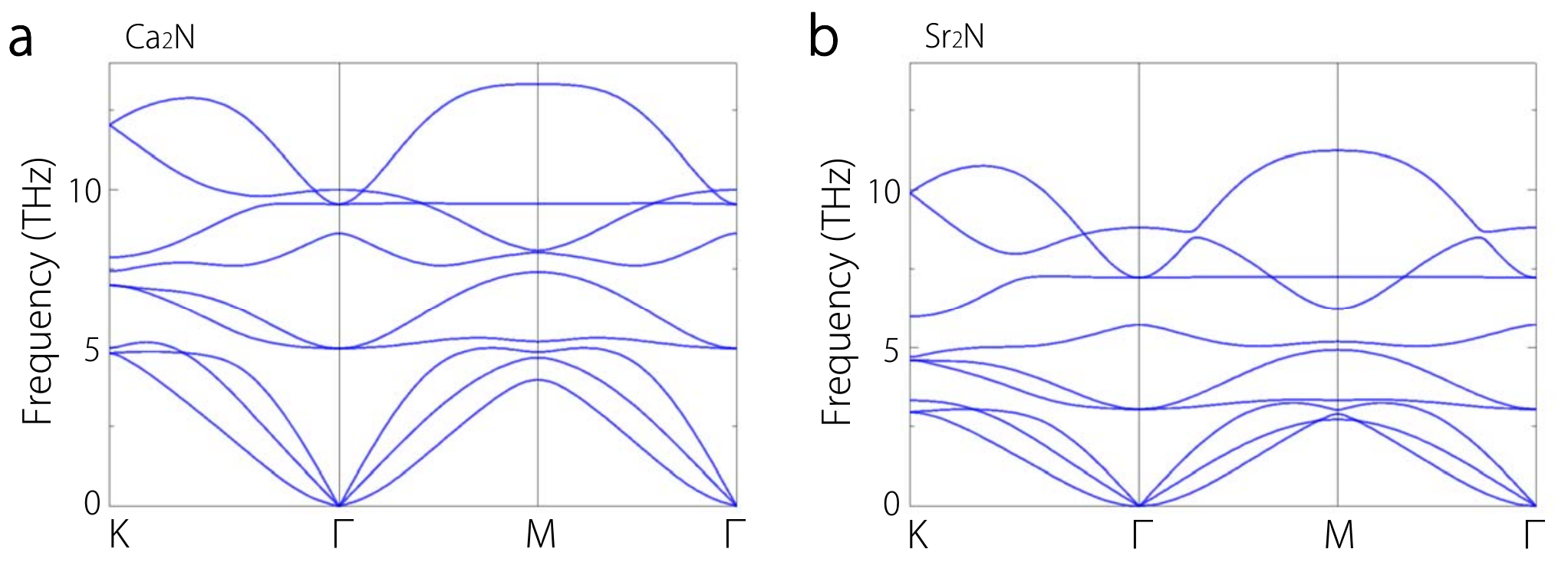,width=16cm}
  \end{center}
  \caption{\label{Fig2} \textbf{Phonon dispersions}. \textbf{a}, Phonon dispersion of 1-ML Ca$_2$N. \textbf{b}, Phonon dispersion of 1-ML Sr$_2$N. }
\end{figure}

The dynamical stability of 1-ML Ca$_2$N has been studied in Ref.\onlinecite{Zhao2014}. Here we find that few-layer Ca$_2$N and Sr$_2$N (including Sr$_2$N monolayer) are also dynamically stable through analysis of their phonon spectra. In contrast, the similar structures of 1-ML and 2-ML Ba$_2$N are not stable due to the presence of imaginary frequencies in their phonon spectra. Therefore, we only focus on Ca$_2$N and Sr$_2$N in the present work. As representative examples, the phonon dispersions of monolayer Ca$_2$N and Sr$_2$N are shown in Fig.\ref{Fig2}. The absence of imaginary frequencies in the whole Brillouin zone demonstrates that the dynamical stability of the corresponding structures. The in-plane and out-of-plane transverse acoustic modes are not degenerate due to the structural anisotropy which is typical for layered materials. More importantly, while the in-plane transverse acoustic modes have a linearly dependence on their wavevectors in the vicinity of $\Gamma$-point, the out-of-plane acoustic (ZA) phonons exhibit a quadratic dispersion around $\Gamma$-point. The parabolic dispersion of ZA modes is a characteristic feature of layered materials, which is consistent with the macroscopic elastic theory of thin plates\cite{Zabel2001}. This feature has been frequently observed in other layered 2D materials, such as graphene\cite{Liu2007a}, layered transition metal dichalcogenides\cite{Sanchez2011}, and phosphorene\cite{Zhu2014}.

\begin{figure}
  \begin{center}
   \epsfig{file=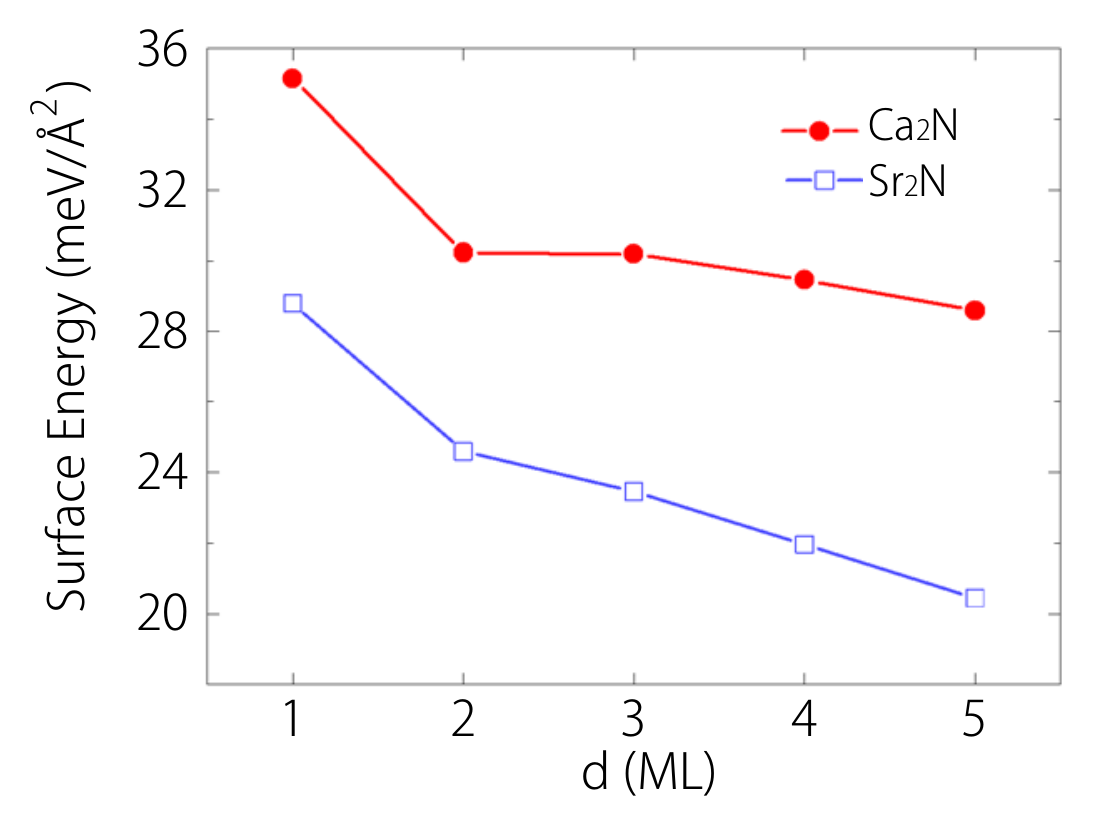,width=10cm}
  \end{center}
  \caption{\label{Fig3} Surface energies for few-layer Ca$_2$N and Sr$_2$N as a function of thickness $d$ from 1-ML to 5-ML.}
\end{figure}

We then calculate the surface energies of 1-ML to 5-ML Ca$_2$N and Sr$_2$N and the results are plotted in Fig.\ref{Fig3}. Because of their layered structure and the large interlayer separations, the surface energies are relatively low. The calculated values are comparable to the surface energy of graphene ($\sim12$meV/\textup{\AA}$^2$)\cite{Zacharia2004}. Previous calculations on 1-ML Ca$_2$N have suggested its possibility to be mechanically exfoliated from the bulk. From our result in Fig.\ref{Fig3}, the surface energy decreases with the increasing number of layers. Hence few-layer Ca$_2$N would be more easily exfoliated compared with monolayer Ca$_2$N, as naturally expected. Sr$_2$N has an even larger interlayer spacing.  Surface energy for 1-ML Sr$_2$N is about $29$meV/\textup{\AA}$^2$, and it decreases to about $20$meV/\textup{\AA}$^2$ for 5-ML. With the same number of layers, Sr$_2$N's surface energy is smaller than that of Ca$_2$N by about $6\sim8$meV/\textup{\AA}$^2$. This implies that Sr$_2$N monolayer and few-layers could be more easily obtained by mechanical exfoliation using Scotch tapes or AFM tips.

\subsection{Electronic structure and 2D confined electron layers.}
\begin{figure}
  \begin{center}
   \epsfig{file=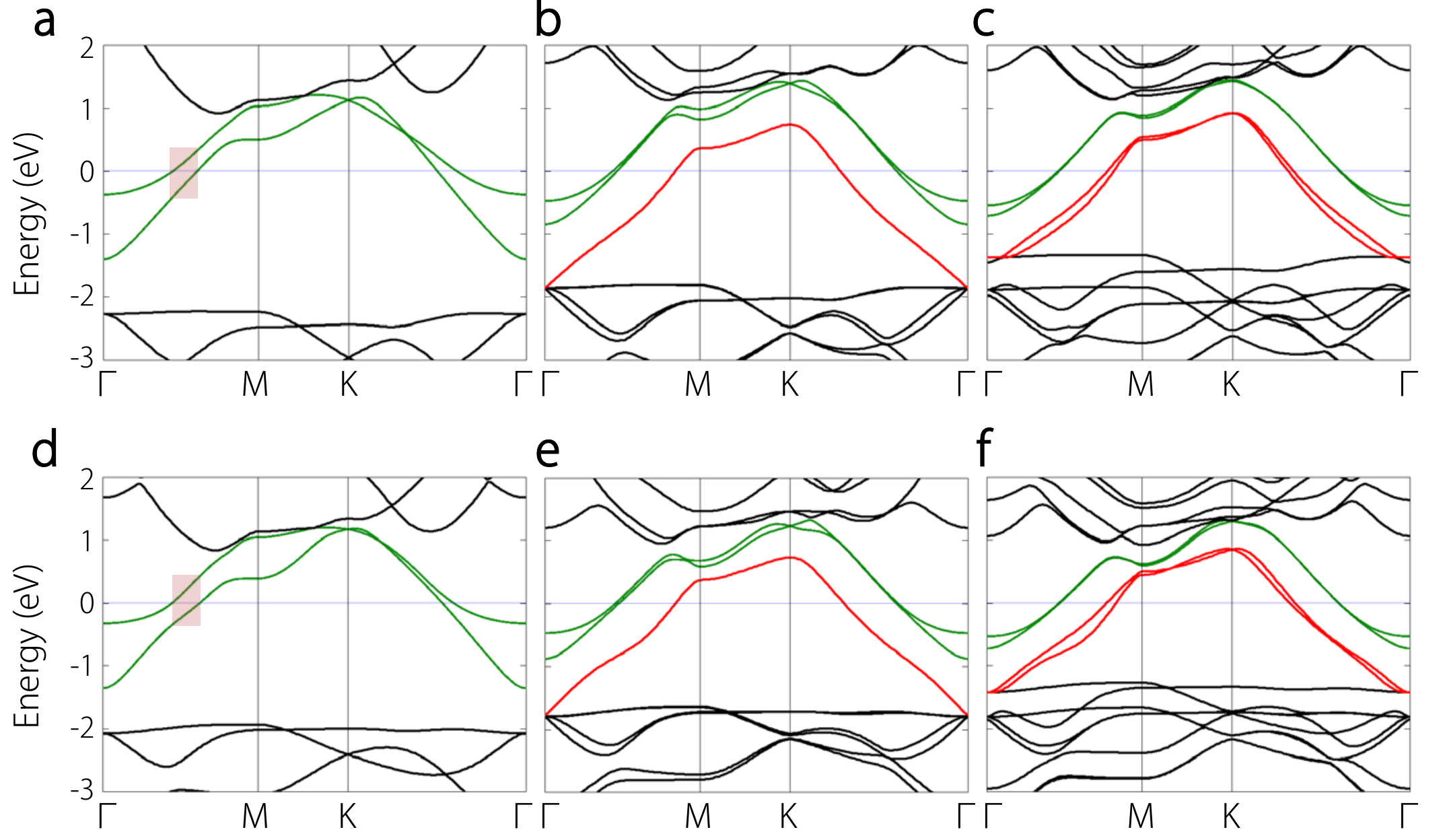,width=16cm}
  \end{center}
  \caption{\label{Fig4} \textbf{Electronic band structures of few-layer Ca$_2$N and Sr$_2$N}. \textbf{a-c}, Band structure of Ca$_2$N with \textbf{a} 1-ML, \textbf{b} 2-ML, and \textbf{c}, 3-ML thickness. \textbf{d-f}, Band structure of Sr$_2$N with \textbf{d} 1-ML, \textbf{e} 2-ML, and \textbf{f}, 3-ML thickness. Fermi energy is set at zero. The two green colored bands are mainly from the 2D electron layers confined to the surface. The red colored bands (for films thicker than 1-ML) are mainly from the 2D electron layers confined in the interlayer regions. The red shaded rectangles in \textbf{a} and \textbf{d} indicate the regions contribute to the peaks $\sim$0.3eV in interband $\mathrm{Im}\varepsilon(\omega)$ for 1-ML Ca$_2$N and Sr$_2$N, as observed in Fig4.(a) and Fig.(d). }
\end{figure}

Bulk Ca$_2$N and Sr$_2$N have conduction electrons confined in the 2D interlayer regions playing the role of anions\cite{Lee2013,Walsh2013}. One key question is whether these anionic electron layers are maintained in their thin film structures. We find that the answer is positive. In Fig.\ref{Fig4}, we show the electronic band structures of X$_2$N (X=Ca, Sr) from 1-ML to 3-ML. First, one notices that all these systems are metallic, with Fermi level lying in partially filled dispersive energy bands. In 1-ML Ca$_2$N (and Sr$_2$N), there are two bands crossing Fermi level (see Fig.\ref{Fig4}a and \ref{Fig4}d). By inspecting the charge density distribution, one can verify that these two bands are from the 2D confined electron layers residing on the two sides of the [Ca$_2$N]$^+$ layer, and were referred to as the 2D electron gas in free space states\cite{Zhao2014}. The two bands are energy splitted due to the coupling between the two sides. When a second [Ca$_2$N] layer is added, i.e. for a 2-ML structure, there appears an additional band crossing Fermi level, which has similar dispersion as the two surface bands but is lower in energy (see Fig.\ref{Fig4}b). This band is from the confined anionic electron layer between the two [Ca$_2$N]$^+$ layers, as we will show in the following. The energy for this 2D band is lower because the confinement in the interlayer region is stronger than that for the surface, leading to a deeper potential well. One also observes that the splitting between the two surface bands is reduced, as a result of the reduced coupling between the two with increasing structural thickness. When going to 3-ML, there are two interlayer gap regions and indeed there is one more band crossing Fermi level (see Fig.\ref{Fig4}c). One observes that the two bands from the interlayer confined 2D electrons almost coincide, indicating that these states are strongly confined and coupling between neighboring interlayer regions is very small. This clearly demonstrates the 2D character of the confinement topology. The above features persist when more [Ca$_2$N] layers are added. Then more interlayer confined anionic electron bands will appear and are nearly degenerate in energy. Their dispersion and bandwidth are almost independent on the thickness. For Sr$_2$N, the band structure is quite similar to that of Ca$_2$N with the same number of layers (with a slight decrease of the bandwidth, c.f. Fig.\ref{Fig4}d-f), implying that the 2D electride character is also maintained in its thin film form.

\begin{figure}
  \begin{center}
   \epsfig{file=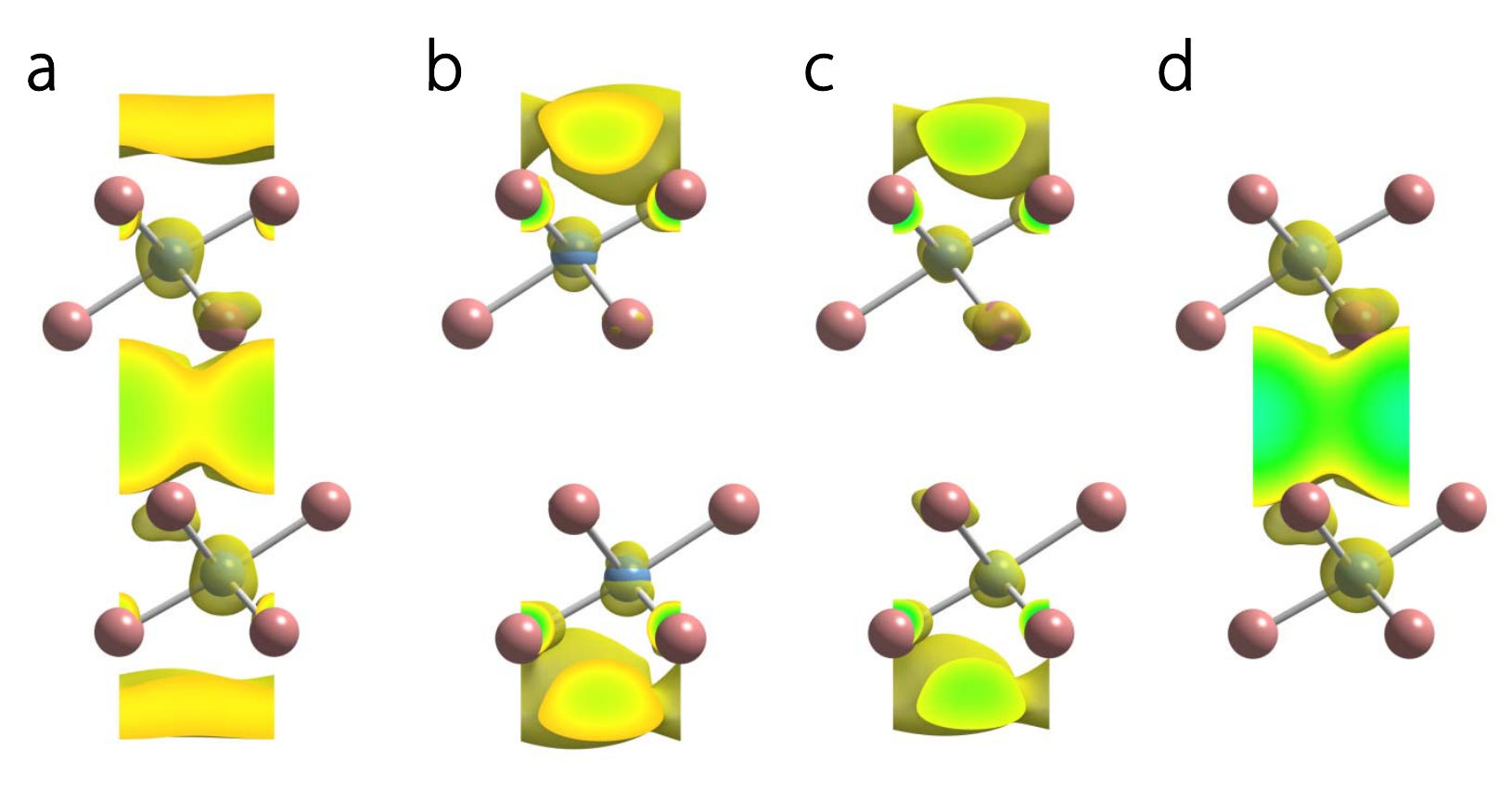,width=12cm}
  \end{center}
  \caption{\label{Fig5} \textbf{Electron density distribution in 2-ML Ca$_2$N}. \textbf{a}, Partial electron density isosurfaces (with value of 0.0003/Bohr$^3$) for states in the energy range $|E-E_f|<$0.05eV shown for a conventional unit cell. \textbf{b-d}, Band-decomposed electron density isosurfaces (with value of 0.003/Bohr$^3$) for the three (colored) bands which cross Fermi level as shown in Fig.\ref{Fig4}(b). \textbf{b} is for the upper green band, \textbf{c} is for the lower green band, and \textbf{d} is for the red band. }
\end{figure}

To visualize the real space distribution of the electronic states, we take 2-ML Ca$_2$N as an example, and plot their partial electron densities for states within 0.05eV around Fermi energy $E_F$ in Fig.\ref{Fig5}a. It is clear that the states around Fermi level are mainly located in three confined 2D regions: outside the two surfaces and in the interlayer space between the [Ca$_2$N]$^+$ layers. We further plot the total charge densities of each of the three conduction bands for 2-ML Ca$_2$N. As shown in Fig.\ref{Fig5}b-d, the lower band is from the 2D states in between the [Ca$_2$N]$^+$ layers, while the two higher bands are from the states confined at the surfaces. Similar result holds for Sr$_2$N. This analysis validates our claims in the previous discussion.

\begin{figure}
  \begin{center}
   \epsfig{file=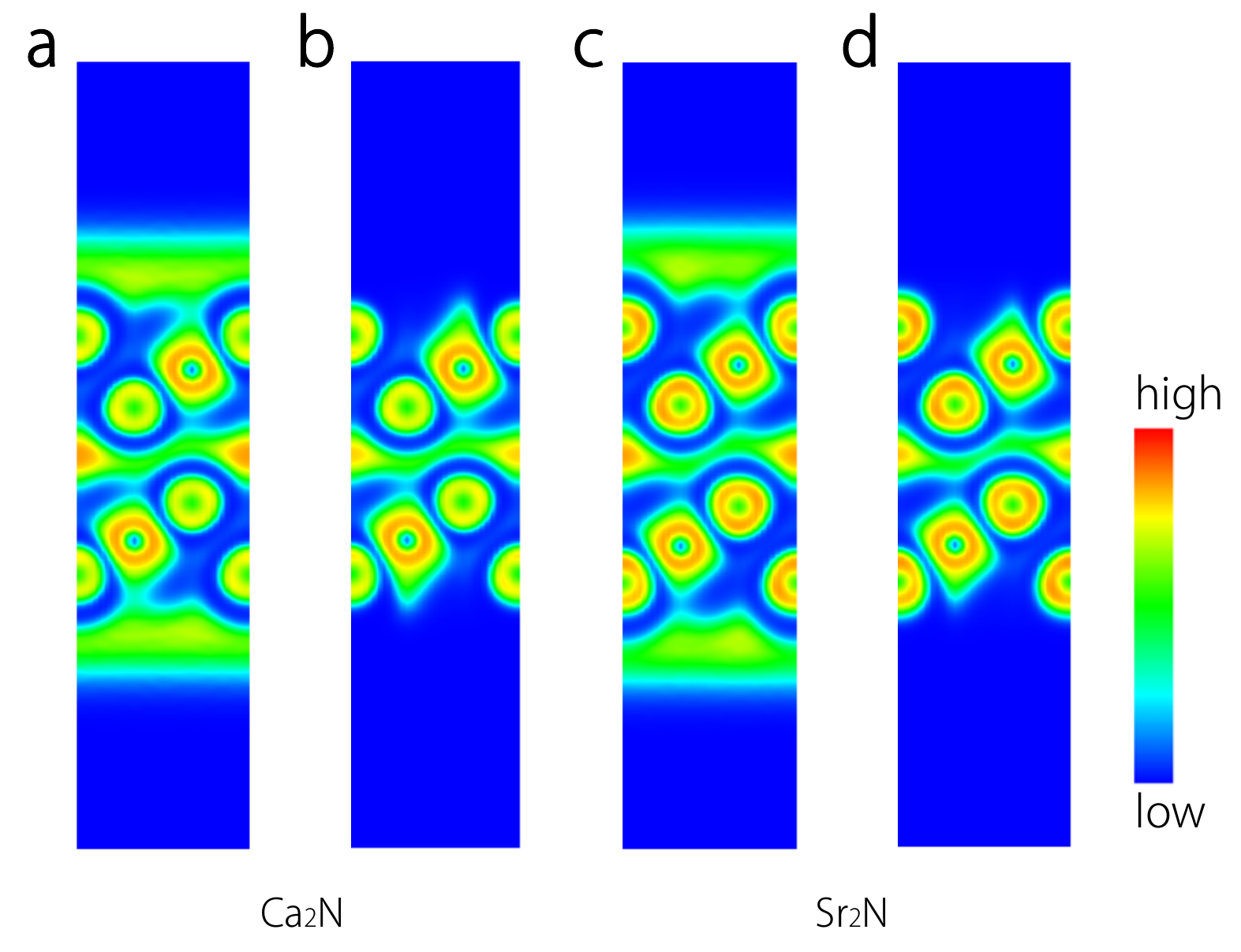,width=10cm}
  \end{center}
  \caption{\label{Fig6} \textbf{Electron localization function (ELF) maps for 2-ML Ca$_2$N and Sr$_2$N}. \textbf{a} and \textbf{c} are the ELF maps for Ca$_2$N and Sr$_2$N respectively, shown for a conventional unit cell. \textbf{b} and \textbf{d} are for [Ca$_2$N]$^+$ and [Sr$_2$N]$^+$ respectively, where one valence electron is removed. }
\end{figure}

Electron-localization function (ELF) is useful for the analysis of the degree of electron localization and the bonding character\cite{Becke1990,Silvi1994}. In Fig.\ref{Fig6}a and \ref{Fig6}c, we show the ELF maps of 2-ML Ca$_2$N and 2-ML Sr$_2$N. Three delocalized electron layers can be seen. One observes that there is no bonding localization attractor between the confined electron layers and the [Ca$_2$N]$^+$ ([Sr$_2$N]$^+$) layers, indicating the bonding between them is of ionic type. When one valence electron is removed, the surface confined layers will be vacated while the interlayer anionic electron layer remain largely intact, as shown in Fig.\ref{Fig6}b and \ref{Fig6}d. This feature also implies that the out-of-plane work function of these layered materials are mainly determined by the 2D electron layers on the surface.

\begin{figure}
  \begin{center}
   \epsfig{file=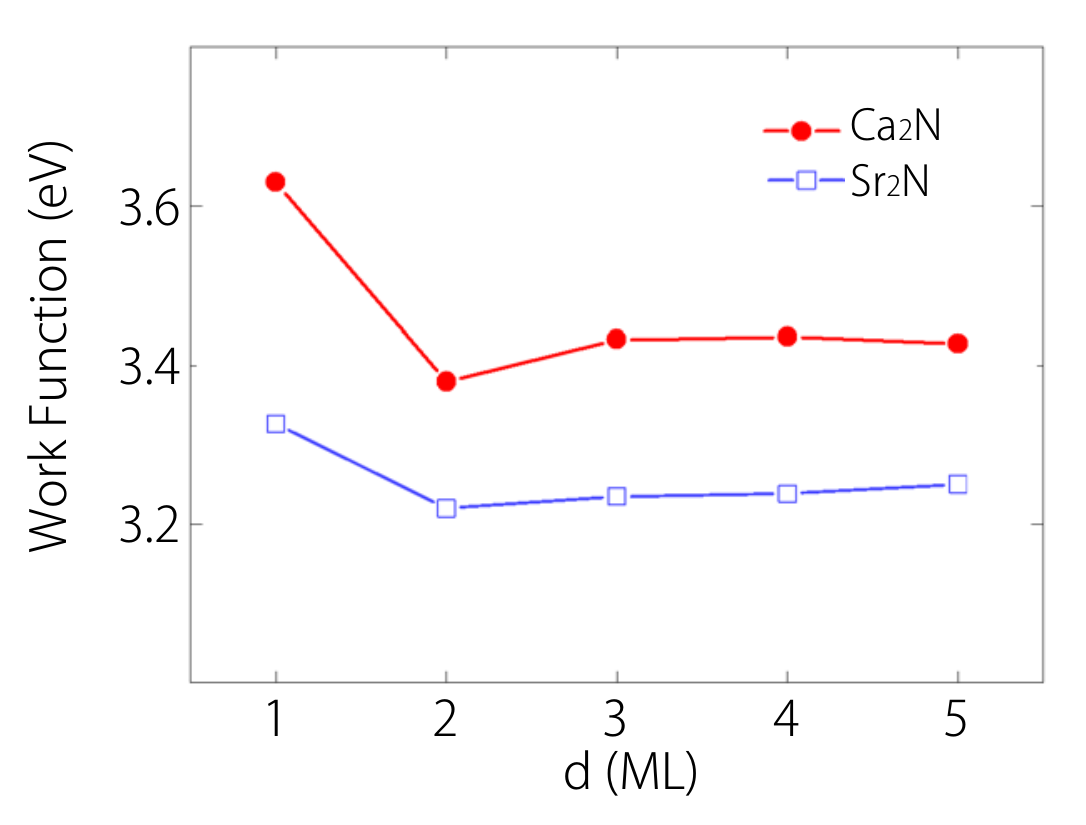,width=10cm}
  \end{center}
  \caption{\label{Fig7} Work functions for few-layer Ca$_2$N and Sr$_2$N as a function of thickness $d$ from 1-ML to 5-ML.}
\end{figure}

\begin{figure}
  \begin{center}
   \epsfig{file=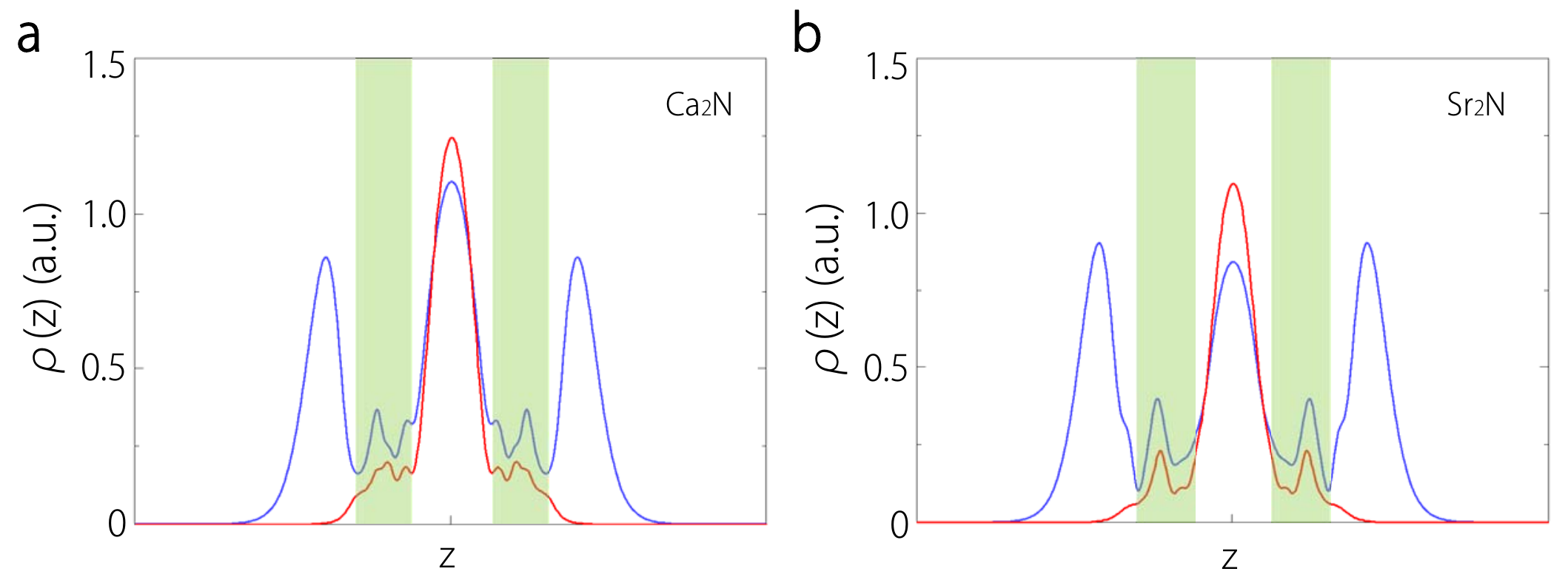,width=14cm}
  \end{center}
  \caption{\label{Fig8} Partial density of states for the energy range $|E-E_f|<$0.05eV averaged over the $ab$ plane for \textbf{a} 2-ML Ca$_2$N and \textbf{b} 2-ML Sr$_2$N. In each figure, the blue curve and the red curve are for the distributions before and after one valence electron is removed, respectively. The green shaded regions indicate the locations of the [Ca$_2$N]$^+$ or [Sr$_2$N]$^+$ layers.}
\end{figure}

Previous studies found that bulk Ca$_2$N has a highly anisotropic work function ranging from 2.6eV (for (100) surface) to 3.4eV (for (001) surface)\cite{Uijitewaal2004,Lee2013}. In Fig.\ref{Fig7}, we plot our results of the out-of-plane work function as a function of the number of layers for both Ca$_2$N and Sr$_2$N. One notes that there is a sharp decrease of the work function from 1-ML to 2-ML for both materials, due to the appearance of the strongly confined interlayer states that decouples the two surfaces. For Ca$_2$N, with increasing layer number, the work function approaches a value around 3.4eV, which is consistent with previous studies. The work function of Sr$_2$N is less than that of Ca$_2$N by about 0.2eV, showing that the electrons in Sr$_2$N are more loosely bound. We average the partial electron density around Fermi level in the layer plane. The resulting 1D density profiles for 2-ML with and without one electron removed are plotted in Fig.\ref{Fig8}. One observes that there are three peaks corresponding to the three anionic electron layers. After one electron is taken away, the two side peaks disappear while the central peak remains. The two side peaks of Sr$_2$N have larger width than those of Ca$_2$N, implying a higher degree of delocalization of these surface states hence a lower work function.

\subsection{Dielectric function.}
The first-principles calculation based on DFT has proved to be a powerful tool for the study of dielectric function for metals, including ultrathin metallic films down to few-layer thickness\cite{Harl2007,Harl2008,He2010,Glantschnig2010,Yan2011,Laref2013,Ming2014}. In crystalline solids, the dielectric function $\varepsilon(\omega)$ consists of two contributions: a Drude-like intraband contribution and an interband contribution. The imaginary part of the interband contribution involves the interband matrix elements of the momentum operators, and can be evaluated directly in DFT\cite{Harl2008}. Its real part can then be calculated via the Kramers-Kronig relation. The intraband contribution is typically treated by the Drude model, in which the plasma frequency (tensor) $\omega_p$ can be evaluated by DFT\cite{Harl2008}. The calculation details are presented in the Method section.

\begin{figure}
  \begin{center}
   \epsfig{file=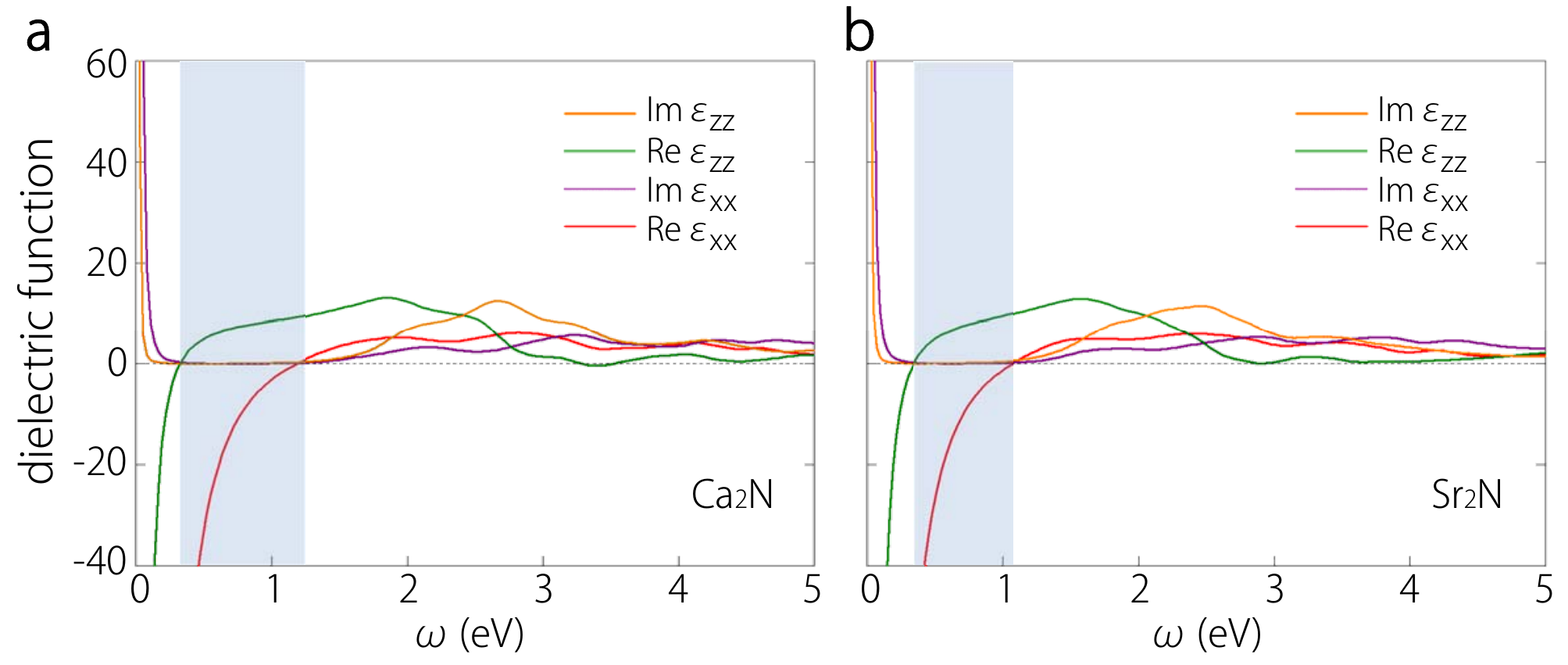,width=14cm}
  \end{center}
  \caption{\label{Fig9} \textbf{Anisotropic dielectric functions for bulk X$_2$N}: \textbf{a} for Ca$_2$N and \textbf{b} for Sr$_2$N. The real and imaginary parts of in-plane component $\varepsilon_{xx}$ and out-of-plane component $\varepsilon_{zz}$ are plotted using different colors. The blue shaded region in each figure indicates the frequency range in which $\mathrm{Re}\varepsilon_{xx}$ and $\mathrm{Re}\varepsilon_{zz}$ have different signs.}
\end{figure}

We first consider the dielectric properties of bulk Ca$_2$N and Sr$_2$N. Simple metals such as Au, Ag, and Al generally have a large isotropic bulk plasma frequency around 10eV. In contrast, Ca$_2$N and Sr$_2$N are of layered structure, hence their dielectric properties are expected to exhibit intrinsic anisotropy. We find that bulk Ca$_2$N has an in-plane (i.e. in the layer plane) plasma frequency $\omega_{p,xx}$ of 3.14eV and an out-of-plane plasma frequency $\omega_{p,zz}$ of 0.95eV; bulk Sr$_2$N has an in-plane plasma frequency of 2.94eV and an out-of-plane plasma frequency of 1.05eV. These frequencies are significantly lower than those for simple metals, partly because the conducting electron density is lower. For each material, the out-of-plane plasma frequency is much lower compared with the in-plane plasma frequency, showing that the neighboring anionic electron layers are only weakly coupled. In Fig.\ref{Fig9}, we plot the real and imaginary part of in-plane and out-of-plane dielectric functions for the two materials. They show the character of metallic behavior, i.e. Drude peaks at low energy due to intraband contribution and the real part of $\varepsilon(\omega)$ crossing from negative value to positive value with increasing frequency. The strong anisotropy in the dielectric function is also quite obvious. Remarkably, because of this anisotropy, the real part of in-plane dielectric function, $\textrm{Re}\varepsilon_{xx}(\omega)$, changes sign at a frequency which is different from that for $\textrm{Re}\varepsilon_{zz}(\omega)$, leading to an extended frequency window in which the two components have different signs (marked as the blue shaded regions in Fig.\ref{Fig9}). This sign difference is the characteristic feature of so-called indefinite media\cite{Smith2003}, which was proposed in the study of metamaterials and has important potential applications such as near-field focusing and building hyperlenses that can transform evanescent fields into propagating modes\cite{Jacob2006,Liu2007}. Previous realizations of indefinite media are mostly in artificially assembled structures which require complicated fabrication process and usually have high dissipation. Our results suggest that crystalline solids Ca$_2$N and Sr$_2$N in their bulk form would just be indefinite materials for a frequency range spanning the near infrared. Moreover, one notes that the imaginary parts of $\varepsilon_{xx}$ and $\varepsilon_{zz}$ in this frequency range are very small ($\sim 0.1$), implying that they could be ideal low-loss indefinite materials.

\begin{figure}
  \begin{center}
   \epsfig{file=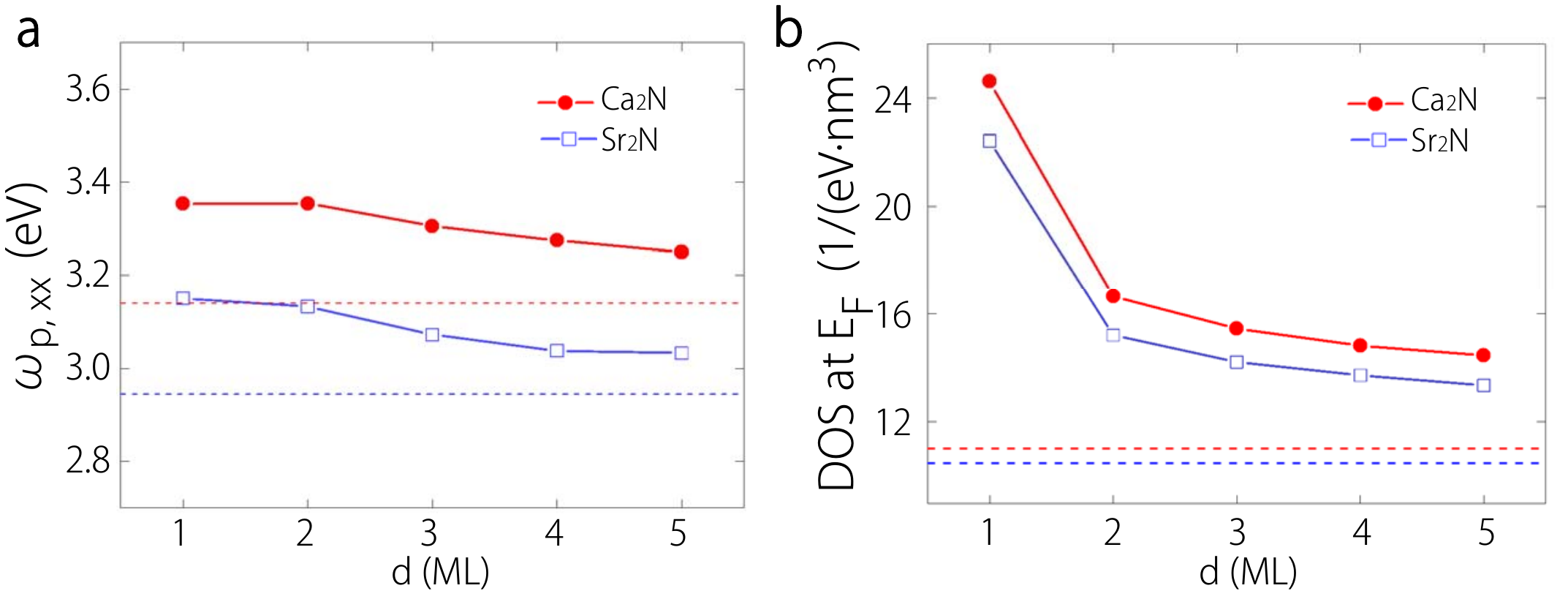,width=14cm}
  \end{center}
  \caption{\label{Fig10} \textbf{a}, In-plane plasma frequency $\omega_{p,xx}$ for Ca$_2$N and Sr$_2$N thin films as a function of thickness $d$ from 1-ML to 5-ML. \textbf{b}, electron density of states (DOS) of Ca$_2$N and Sr$_2$N at Fermi energy as a function of film thickness. In each figure, the dashed lines indicate the corresponding values for the bulk.}
\end{figure}

Next we focus on the dielectric functions of Ca$_2$N and Sr$_2$N few-layer structures. With decreasing film thickness, one expects that the anisotropy effects would be even stronger. The intraband contribution to the out-of-plane dielectric response $\varepsilon_{zz}$ becomes negligible, and it has been shown that for simple metals which are isotropic in bulk can exhibit strongly anisotropic $\varepsilon(\omega)$ down to few-layer thickness\cite{Laref2013}. In Fig.\ref{Fig10}a, we plot the results of in-plane plasma frequency $\omega_{p,xx}$ as a function of the number of layers for both materials. One observes that the frequency slowly decreases with increasing number of layers and approaches its bulk value. The plasma frequency of Sr$_2$N is lower than that of Ca$_2$N by about 0.2eV for each thickness. The square of plasma frequency is roughly proportional to the product of the carrier density and the inverse of effective mass. In Fig.\ref{Fig10}b, one can see that from 1-ML to 2-ML, there is in fact a sharp decrease in the carrier density, as characterized by the DOS per unit volume around Fermi level, primarily due to the appearance of the large interlayer space of 2-ML. However, from the band structure in Fig.\ref{Fig4}b, one notices that the additional 2D electron band (marked in red color) has a smaller effective mass around Fermi level than the two surface bands, hence decreasing the average effective mass. This compensates the decrease in carrier density, resulting in an overall small change in the plasma frequency between 1-ML and 2-ML.

\begin{figure}
  \begin{center}
   \epsfig{file=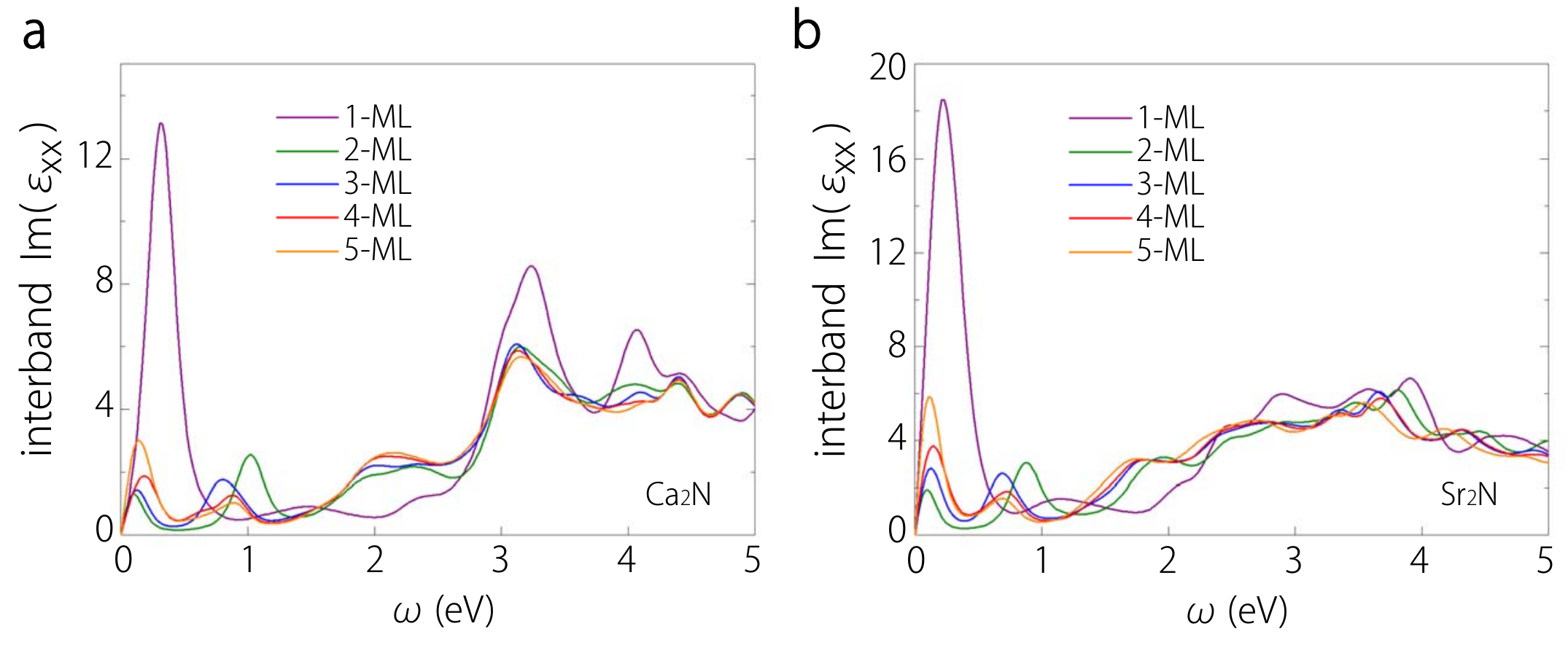,width=14cm}
  \end{center}
  \caption{\label{Fig11} The interband contribution to $\mathrm{Im}\varepsilon_{xx}(\omega)$ for \textbf{a} Ca$_2$N and \textbf{b} Sr$_2$N thin film structures from 1-ML to 5-ML. Results for different thicknesses are plotted using different colors.}
\end{figure}

In Fig.\ref{Fig11}, we show the interband contribution to the imaginary part of the in-plane dielectric function, $\textrm{Im}\varepsilon_{xx}(\omega)$, which is closely connected to band structure features. One noticeable feature is the sharp peak around $0.3$eV for both materials at 1-ML thickness. This peak has a sharp drop from 1-ML to 2-ML and shifts to a lower energy. Its appearance can be attributed to the (almost) parallel sections of the two surface bands for 1-ML, as indicated in the red shaded regions in Fig.\ref{Fig4}a and \ref{Fig4}d. From 1-ML to 2-ML, the splitting between two surface bands decreases and the volume increases a lot due to the appearance of the large interlayer region, resulting in the observed change of the peak. For thicknesses larger than 1-ML, there is another peak around 1eV, which is due to the transition between the interlayer 2D bands and the surface bands (c.f. Fig.\ref{Fig4}). From the band structures, it is clear that the response below 1eV is mainly contributed by the states from the 2D anionic electron layers.

\begin{figure}
  \begin{center}
   \epsfig{file=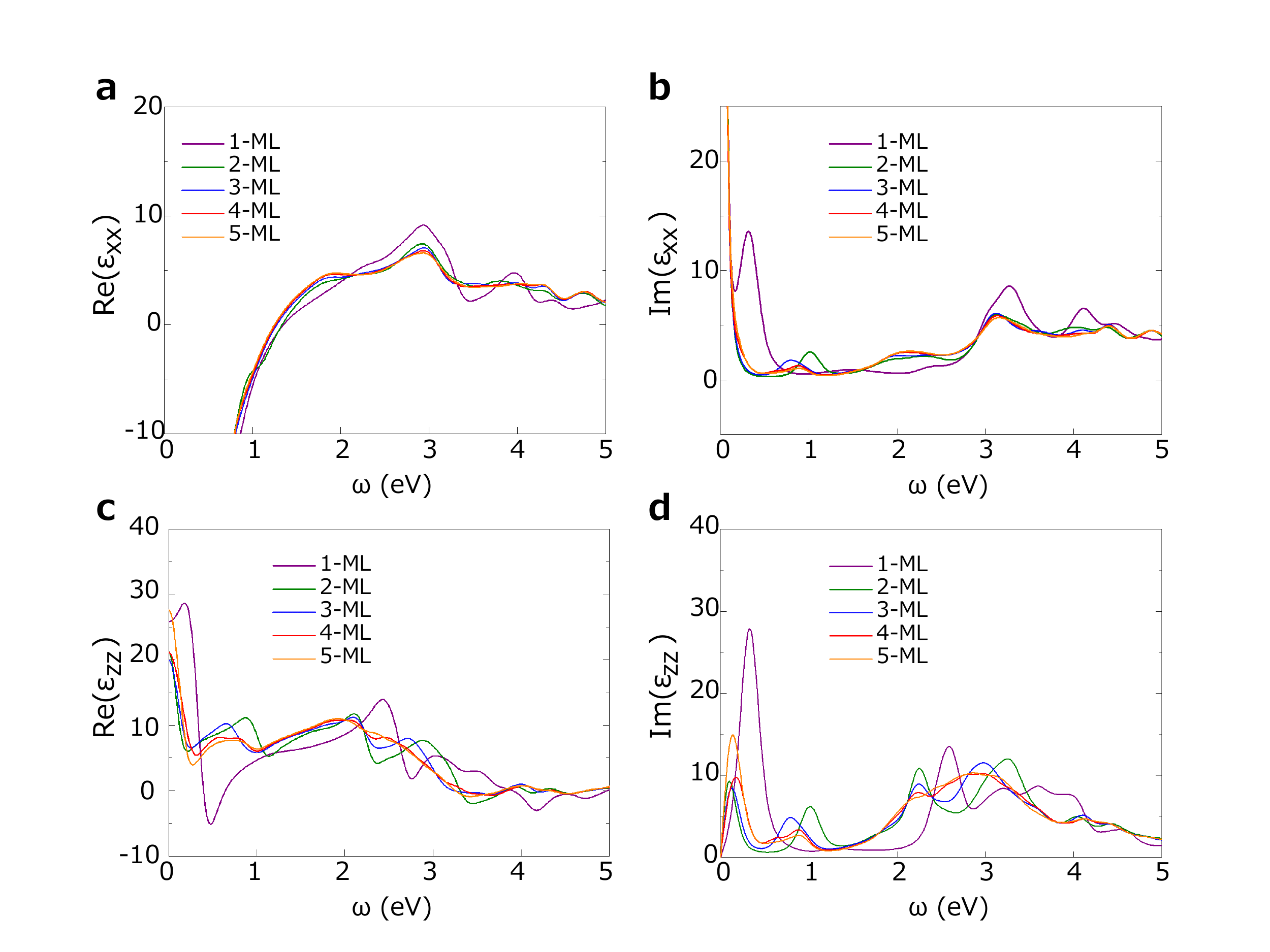,width=14cm}
  \end{center}
  \caption{\label{Fig12} \textbf{Dielectric functions for Ca$_2$N few-layers.} \textbf{a} and \textbf{b} are for the real and imaginary parts of in-plane component $\varepsilon_{xx}$ respectively. \textbf{c} and \textbf{d} are for the real and imaginary parts of out-of-plane component $\varepsilon_{zz}$ respectively. Results for different thicknesses are plotted using different colors.}
\end{figure}

\begin{figure}
  \begin{center}
   \epsfig{file=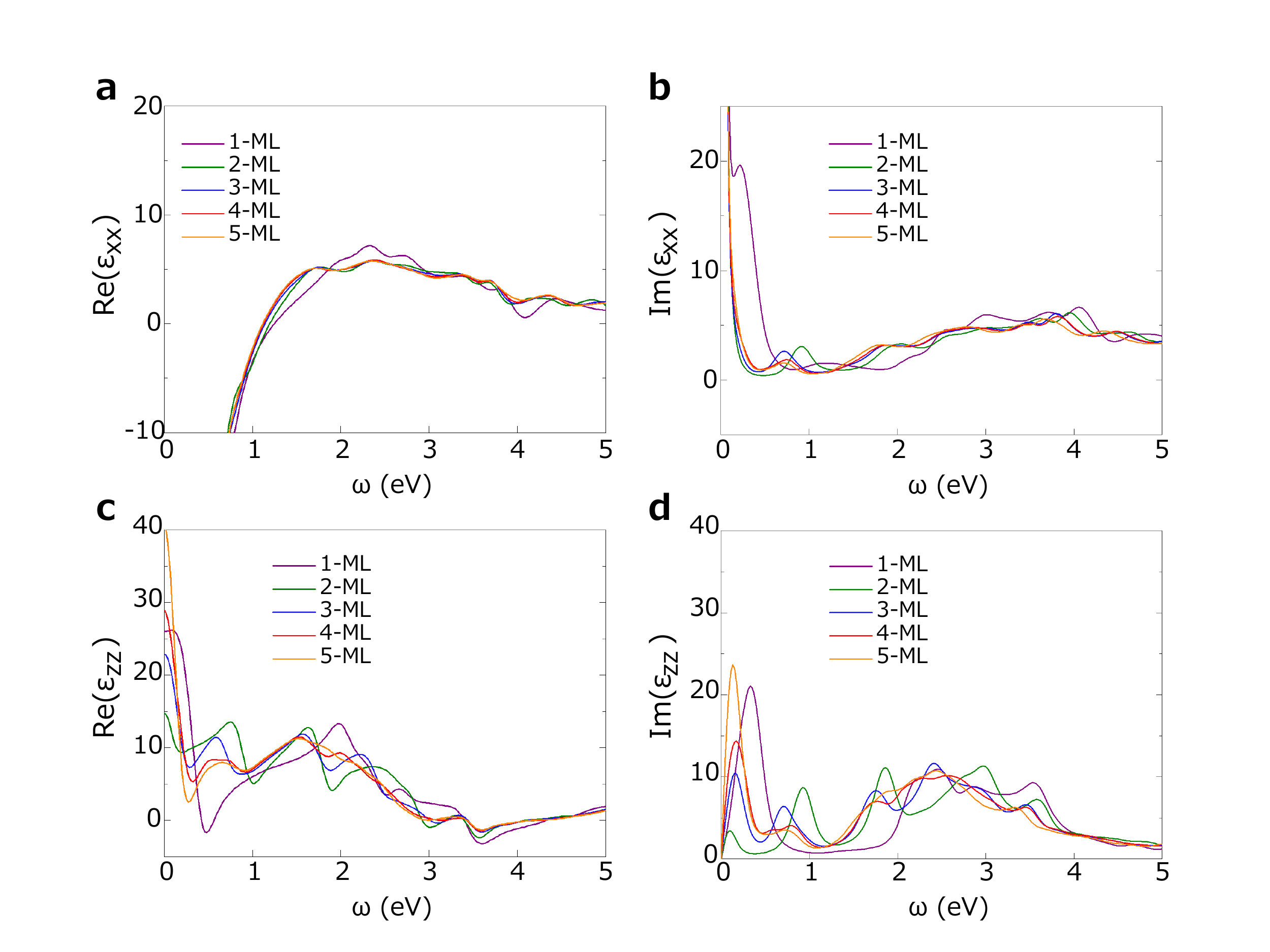,width=14cm}
  \end{center}
  \caption{\label{Fig13} \textbf{Dielectric functions for Sr$_2$N few-layers.} \textbf{a} and \textbf{b} are for the real and imaginary parts of in-plane component $\varepsilon_{xx}$ respectively. \textbf{c} and \textbf{d} are for the real and imaginary parts of out-of-plane component $\varepsilon_{zz}$ respectively. Results for different thicknesses are plotted using different colors.}
\end{figure}

The total in-plane and out-of-plane dielectric function with their real and imaginary parts for Ca$_2$N few-layers are plotted in Fig.\ref{Fig12}. For $\varepsilon_{xx}$, the low energy part is dominated by the intraband Drude-like contribution. $\textrm{Re}\varepsilon_{xx}(\omega)$ is negative below about 1eV. Meanwhile, the intraband contribution is negligible for $\varepsilon_{zz}(\omega)$. Hence $\textrm{Re}\varepsilon_{zz}(\omega)$ is largely positive for the low energy part. The sign difference between $\textrm{Re}\varepsilon_{zz}$ and $\textrm{Re}\varepsilon_{xx}$ in the low energy range again signals a possible indefinite material. This mechanism for the anisotropic dielectric function has been discussed in the case of Au ultrathin films\cite{Laref2013}. One observes that the change of $\varepsilon(\omega)$ against thickness is most dramatic from 1-ML to 2-ML, due to the introduction of the first interlayer anionic electron layer. The imaginary part of dielectric function for 1-ML is much larger than that for 2-ML, particularly in the near infrared frequency range, which indicates more energy dissipation for possible plasmon modes as we discuss later.
In addition, the variation of $\varepsilon_{xx}$ versus thickness is overall less dramatic compared with that of $\varepsilon_{zz}$, which again reflects the fact that the conducting electrons are strongly confined within the 2D in-plane layered regions. Due to the same reason that the different conducting 2D layers are only weakly coupled, the results for both in-plane and out-of-plane components converges rapidly with increasing thickness (note that the low energy part of the out-of-plane component would eventually be dominated by a Drude-like intraband contribution when approaching the bulk limit). One notes that there is only small difference between the results of 4-ML and 5-ML. We also calculated 8-ML which is the largest system size within our computing capability. The obtained results are very close to those for 5-ML. Similar features are also demonstrated in the results for Sr$_2$N few-layer structures, as shown in Fig.\ref{Fig13}.

\subsection{Surface plasmon modes.}
Surface plasmon modes are confined electromagnetic excitations propagating at an interface between a conductor and a dielectric\cite{Maierbook}. It typically requires a sign change of $\textrm{Re}\varepsilon$ across the interface. Therefore the most commonly used materials for plasmonic applications are metals such as Au, Ag, and Al which have a range of frequencies with negative $\textrm{Re}\varepsilon(\omega)$\cite{West2010}. Since Ca$_2$N and Sr$_2$N are also metals, one may also wonder whether they could also support surface plasmon modes. This question is particularly interesting when considering the frequency range in which the material shows indefinite medium property. In such case, one expects that the positive $\mathrm{Re}\varepsilon_{zz}$ component may compete with the negative $\mathrm{Re}\varepsilon_{xx}$ component and tend to destroy the bounded plasmon modes.

\begin{figure}
  \begin{center}
   \epsfig{file=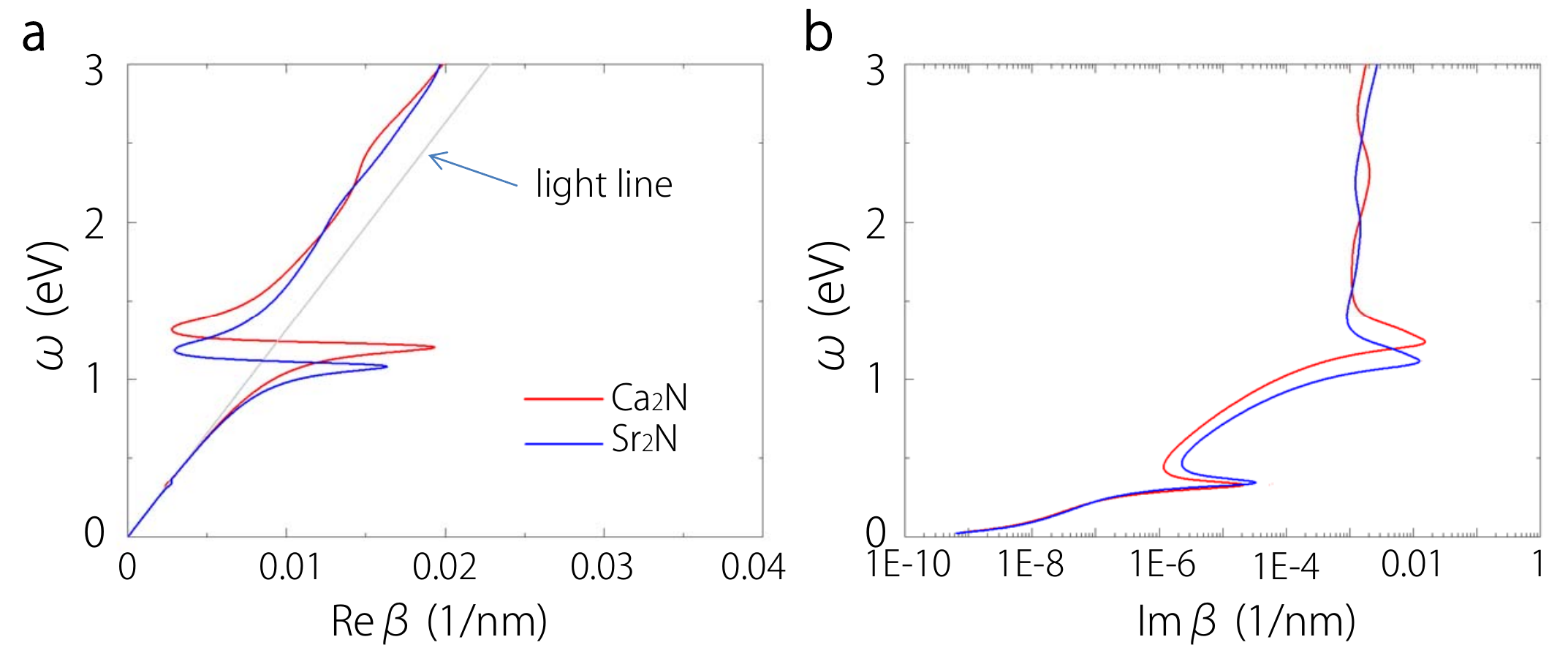,width=14cm}
  \end{center}
  \caption{\label{Fig14} Dispersion relation of surface plasmon modes for an interface between a dielectric medium with $\varepsilon_d=2.25$ and bulk X$_2$N (X=Ca, Sr). The light line in the dielectric medium is also shown in grey color.}
\end{figure}

Let's first consider the interface between bulk Ca$_2$N or Sr$_2$N and a dielectric medium characterized by a frequency-independent dielectric constant $\varepsilon_d>0$. The interface is parallel to the layer plane. As we have shown, Ca$_2$N and Sr$_2$N in the bulk form already have strong anisotropy in their dielectric functions. Assume that the interface supports a transverse magnetic (TM) plasmon mode travelling along the interface with a wave vector $\beta$. Following standard derivation using Maxwell's equations\cite{Maierbook}, one obtains that
\begin{equation}\label{bulkbeta}
\beta=k_0\sqrt{\frac{\varepsilon_d\varepsilon_{zz}(\varepsilon_{xx}-\varepsilon_d)}{\varepsilon_{xx}\varepsilon_{zz}-\varepsilon_d^2}}.
\end{equation}
Here $k_0=\omega/c$ is the wave vector in vacuum, $c$ is the speed of light, and $\varepsilon_{xx}$ and $\varepsilon_{zz}$ are the two components of the dielectric function for our conducting material. If $\varepsilon_{xx}=\varepsilon_{zz}$, the result in Eq.(\ref{bulkbeta}) reduces to the familiar result for isotropic metal\cite{Maierbook}. The dispersion characteristics of the surface plasmon modes for both materials are shown in Fig.\ref{Fig14}, with $\varepsilon_d=2.25$ (appropriate for SiO$_2$). In Fig.\ref{Fig14}a, the modes lying to the right of the light line (in the dielectric medium) are bounded to the surface. One observes that the results show characteristic surface plasmon peaks similar to simple metals and confined plasmon modes still exist within the frequency range where the material shows indefinite medium property. A major difference is that for simple metals the peak position, known as surface plasmon frequency $\omega_\mathrm{sp}$, occurs at higher energies, e.g. $\omega_\mathrm{sp}\sim 3.4$eV for Ag/SiO$_2$ interface\cite{Dionne2005}; while $\omega_\mathrm{sp}$ is much lower here, around 1.1eV-1.2eV in the near infrared range. The $\mathrm{Im}\beta$ shown in Fig.\ref{Fig14}b is connected to the energy damping during the mode propagation. Since high dissipation occurs around $\omega_\mathrm{sp}$, for practical applications, modes with frequencies less than $\omega_\mathrm{sp}$ are used. Here for Ca$_2$N, if we take $\omega=1.02$eV, the corresponding surface plasmon wavelength $\lambda_\mathrm{sp}\simeq656$nm, and we have a long propagation length $L=1/(2\mathrm{Im}\beta)\simeq5.15\mu$m and the decay length in the dielectric is $l_d=1/\mathrm{Re}\sqrt{\beta^2-\varepsilon_d k_0^2}\simeq178$nm. Longer propagation length can be achieved at lower frequencies, e.g. at $\omega=0.64$eV, we can have $L\simeq173\mu$m but the mode confinement is reduced, with $l_d\simeq737$nm. These values are comparable to those for the noble metals (in the visible or UV frequency range) which are the usual building blocks for plasmonics devices. However, the operating frequency here is much lower. From above discussion, we see that despite the intrinsic anisotropy, bulk Ca$_2$N and Sr$_2$N can still support surface plasmons and could be suitable plasmonics materials in the near infrared frequency range.

\begin{figure}
  \begin{center}
   \epsfig{file=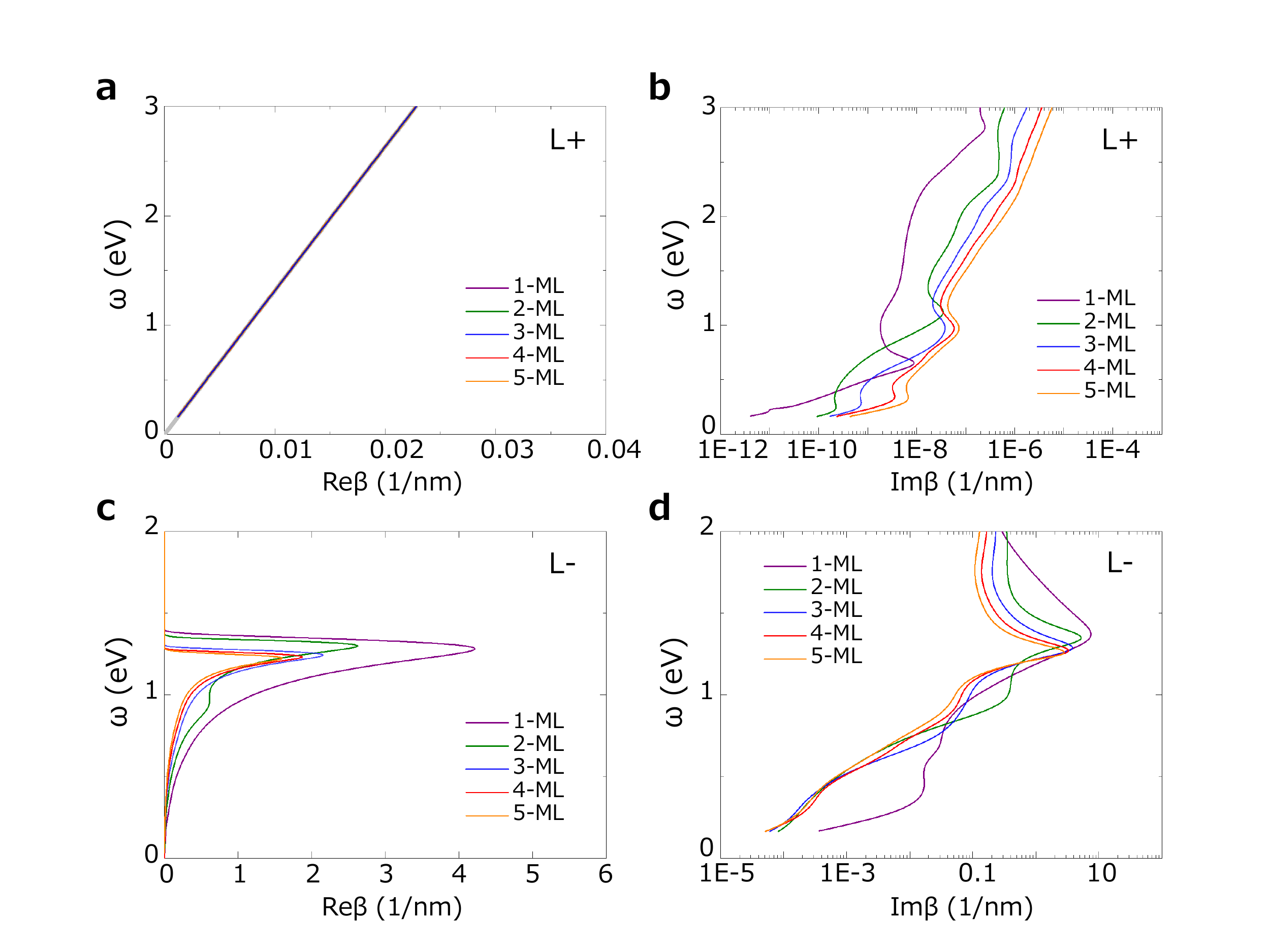,width=14cm}
  \end{center}
  \caption{\label{Fig15} Dispersion characteristics of surface plasmon modes for Ca$_2$N few-layers in a dielectric medium with $\varepsilon_d=2.25$. \textbf{a} and \textbf{b} are for the antisymmetric mode ($L+$ mode). \textbf{c} and \textbf{d} are for the symmetric mode ($L-$ mode). \textbf{a} and \textbf{c} show the real part of the wave number component. \textbf{b} and \textbf{d} show the imaginary part of the wave number component.}
\end{figure}

\begin{figure}
  \begin{center}
   \epsfig{file=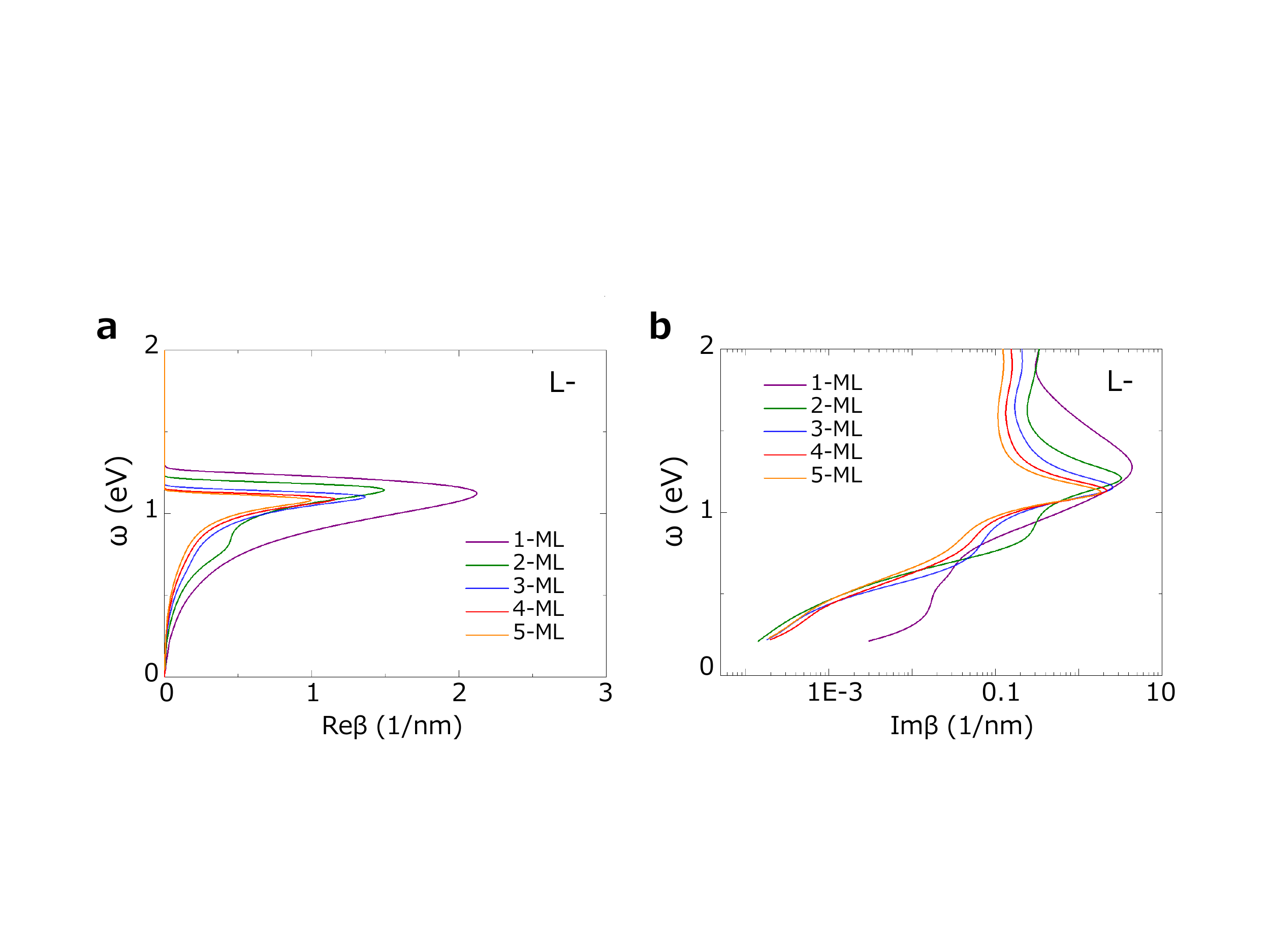,width=14cm}
  \end{center}
  \caption{\label{Fig16} Dispersion characteristics of $L-$ symmetric surface plasmon modes for Sr$_2$N few-layers in a dielectric medium with $\varepsilon_d=2.25$. \textbf{a} shows the real part of the wave number component. \textbf{b} shows the imaginary part of the wave number component.}
\end{figure}

We then turn to the thin films of Ca$_2$N and Sr$_2$N sandwiched between dielectric materials. Besides the change in the thickness-dependent dielectric function, an important effect in thin films is that the surface plasmon modes at two interfaces could couple and form two modes with opposite parity: a symmetric mode ($L-$) and an antisymmetric mode ($L+$)\cite{Dionne2005}. Their dispersions have been derived before in the study of ultrathin metallic films\cite{Laref2013}, and are quoted in our Method section. In Fig.\ref{Fig15}, we plot the surface plasmon dispersion characteristics for Ca$_2$N with different film thicknesses. One observes that the antisymmetric modes are lying on the light line of the dielectric medium, indicating that they are squeezed out of the metal region forming unbounded modes propagating in the dielectric material, which is similar to the case of Au ultrathin films. Meanwhile, pronounced plasmonic peaks do show up for the symmetric modes, clearly indicating that these are bounded surface plasmon modes.
Again, one observes that the variation with thickness is most dramatic between 1-ML and 2-ML. For 1-ML, the bounded modes occur around 1eV. The high peak in $\mathrm{Re}\beta$ vs. $\omega$ shows the modes are strongly bounded to the metallic layer. For thicknesses of 2-ML to 5-ML, the dispersions are quite close. The corresponding surface plasmon frequencies shift to around 1.2eV.
As for the imaginary part of $\beta$, 1-ML structure has relatively large values, whereas $\mathrm{Im}\beta$ for 2-ML to 5-ML almost collapse on a single curve and deceases rapidly in the range below 0.8eV. For Sr$_2$N few-layers, the results for surface plasmon dispersion show similar features. The dispersions for symmetric ($L-$) modes are shown in Fig.\ref{Fig16}. The antisymmetric ($L+$) modes are again unbounded hence are not shown.

\begin{figure}
  \begin{center}
   \epsfig{file=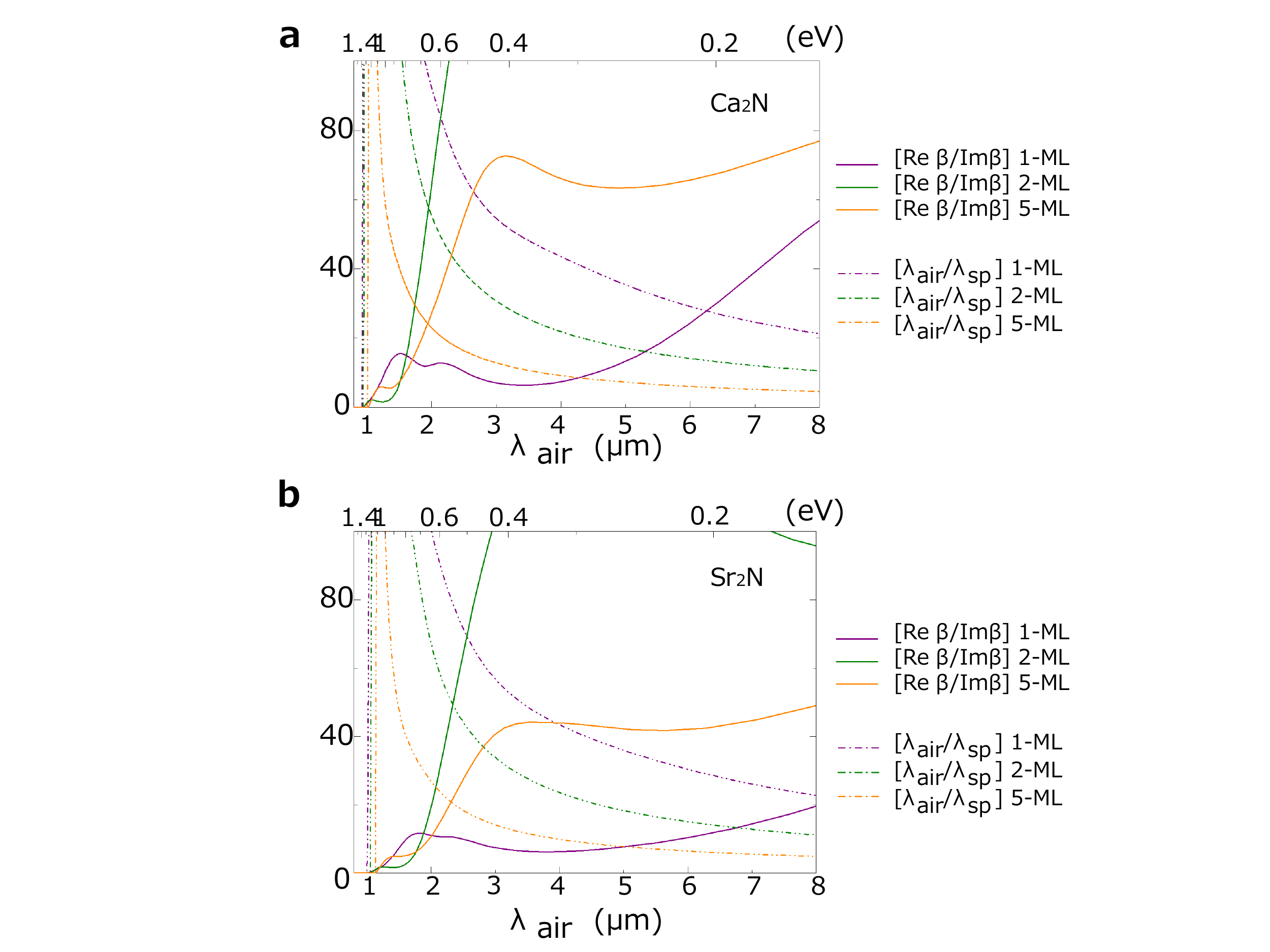,width=14cm}
  \end{center}
  \caption{\label{Fig17} Characteristics $\mathrm{Re}\beta/\mathrm{Im}\beta$ and $\lambda_\mathrm{air}/\lambda_\mathrm{sp}$ for the symmetric surface plasmon modes in few-layer \textbf{a} Ca$_2$N and \textbf{b} Sr$_2$N, plotted as functions of the wavelength in air $\lambda_\mathrm{air}$. $\lambda_\mathrm{sp}$ is the surface plasmon wavelength. The dielectric medium is with $\varepsilon_d=2.25$.}
\end{figure}

Compared with their bulk results, one observes that the surface plasmon frequency $\omega_\mathrm{sp}$ is more or less the same, but both $\mathrm{Re}\beta$ and $\mathrm{Im}\beta$ are greatly increased by two orders of magnitude. This means that the plasmon wavelength and the confinement scale are much smaller, which are desired. However, the propagation length is also decreased at the same time.
Therefore, it is more meaningful to consider the dimensionless ratio $\mathrm{Re}\beta/\mathrm{Im}\beta$, which measures how many surface plasmon wavelengths can be covered before the wave loses most of its energy, as well as the wave localization (or wave shrinkage) quantified by $\lambda_\mathrm{air}/\lambda_\mathrm{sp}$, where $\lambda_\mathrm{air}=2\pi c/\omega$ is the wavelength in air. In Fig.\ref{Fig17}, we show these two dimensionless characteristics for the two materials as functions of $\lambda_\mathrm{air}$. The results for 3-ML and 4-ML are similar to those of 5-ML hence are not shown. One observes that for 1-ML $\mathrm{Re}\beta/\mathrm{Im}\beta$ is lower for most frequencies, due to the relatively high dissipation associated with $\mathrm{Im}\beta$ (c.f. Fig.\ref{Fig15}d). Generally, the wave localization reaches its peak near the surface plasmon resonance. However, there $\mathrm{Re}\beta/\mathrm{Im}\beta$ is small due to the enhanced dissipation. While larger $\mathrm{Re}\beta/\mathrm{Im}\beta$ can be achieved for longer wavelengths, the wave localization becomes poor in that range. This tradeoff is a typical feature for surface plasmons. For application purposes, a compromise can be reached somewhere in between when the two are comparable.

Since 1-ML Ca$_2$N and Sr$_2$N have considerably larger dissipation, in the following we take 2-ML Ca$_2$N and Sr$_2$N as examples. For $\lambda_\mathrm{air}=2\mu$m (0.64eV) in the near infrared range, 2-ML Ca$_2$N has a decay length $l_d=5.4$nm in the dielectric, showing that the plasmon mode is strongly confined to the metallic layer. The ratios $\lambda_\mathrm{air}/\lambda_\mathrm{sp}=56$ and $\mathrm{Re}\beta/\mathrm{Im}\beta=61$. These two values are considerably larger than what noble metals could achieve at their optimal working frequencies. The corresponding values for Sr$_2$N at $\lambda_\mathrm{air}=2\mu$m ($\omega=0.64$eV) are given by $l_d=4.4$nm, $\lambda_\mathrm{air}/\lambda_\mathrm{sp}=68$ and $\mathrm{Re}\beta/\mathrm{Im}\beta=17$. In comparison, the surface plasmon modes for noble metals such as Au and Ag are unbounded at this wavelength. Noticeably, the surface plasmon wavelengths in these materials are very small, $\lambda_\mathrm{sp}=34.9$nm for Ca$_2$N and $\lambda_\mathrm{sp}=28.9$nm for Sr$_2$N (at $\omega=0.64$eV), while the shortest $\lambda_\mathrm{sp}$ for Au and Ag (occur in the visible light range) would be at least larger than 100nm. These suggest that Ca$_2$N and Sr$_2$N thin films have great potential for making plasmonic devices operating in the near infrared range with much smaller scales.

\section*{Discussion}
The loosely bound anionic electrons in electrides are highly reactive in ambient conditions. They could be used as good electron donors and as excellent catalysts for chemical reactions\cite{Kitano2012}. However, for physical applications, one needs to stabilize its property, e.g. by effective encapsulation. For 1-ML Ca$_2$N, a possible encapsulation scheme using 2D insulating layers of graphane was proposed\cite{Zhao2014}.
In the case of few-layer alkaline earth subnitrides, since as we discussed each interlayer anionic electron layer are strongly confined in 2D regions and coupling between layers is small, one expects that the the most reactive electrons are from the surface layers, while the electron layers inside should be less reactive. A detailed study of this point and possible encapsulation schemes for Sr$_2$N will be deferred to a future work.

Compared with Ca$_2$N, Sr$_2$N has a smaller electrostatic potential associated with the larger atomic number of Sr. This was reflected in the lower work function, and may also lead to a high electron mobility for Sr$_2$N few-layers\cite{Walsh2013}. Previous experimental studies have shown that bulk Ca$_2$N has high mobility of 520cm$^2$/(V$\cdot$s)\cite{Lee2013}. One expects that the mobility for Sr$_2$N may be even higher. Hence Sr$_2$N few-layers as new 2D conducting materials could have good potential for electronics applications.

As having been demonstrated in other 2D layered materials, strain engineering has proved to be a powerful tool to modify and control the material properties. The layered materials discussed here have a large interlayer spacing ($>3.5\textup{\AA}$) between the [X$_2$N]$^+$ layers. We also expect that applying strain could be a good method to tune the material properties such as the dielectric function, the plasmonic dispersion, and the carrier mobility. A systematic study of the strain effects is currently underway.

\section*{Methods}
\subsection{First-principles calculations.}
First-principles calculations were carried out using the Vienna \emph{ab}-initio simulation package (VASP)\cite{Kresse1993,Kresse1996}, based on the density functional theory (DFT). The exchange-correlation functional was treated using Perdew-Burke-Ernzerhof generalized gradient approximation\cite{Perdew1996}. The projector augmented wave (PAW) pseudopotential method\cite{Bloechl1994} is employed to model interactions between electrons and ions. The cutoff for plane-wave expansion is set to be 600 eV. The vertical distance between thin films (the thickness of the vacuum gap) is at least 18\textup{\AA}, which is large enough to avoid artificial interactions between the film and its periodic images. Both the atomic positions and lattice constant were fully relaxed using conjugate gradient method. The convergence criteria for energy and force were set to be 10$^{-5}$eV and 0.01eV/\textup{\AA}, respectively. DFT-D2 method was applied to describe the long-range van der Waals interaction. The Brillouin zone integrations have been carried out on a $\Gamma$-centered $k$-mesh. Monkhorst-Pack k-point meshes\cite{Monkhorst1976} with sizes of 15$\times$15$\times$1 and 31$\times$31$\times$1 were used for geometry optimization and static electronic structure calculation, respectively. In the later study of optical properties, the sizes of $k$-mesh are significantly increased to 61$\times$61$\times$1 and 41$\times$41$\times$9 for thin films and bulk respectively, to achieve highly converged results. For the integration over the Brillouin zone in calculating dielectric functions, we used the first order Methfessel-Paxton method\cite{Methfessel989} with a value of 0.1eV. The phonon dispersions of the structures were calculated by using density functional perturbation theory as implemented in PHONOPY code\cite{Togo2008,Gonze1997}.

\subsection{Calculation of dielectric functions.}
The optical properties of solids are mainly due to the response of the electron system to a time-dependent electromagnetic perturbation. For metals, the optical complex dielectric function consists of interband and Drude-like intraband contributions:
\begin{equation}
\varepsilon(\omega)=\varepsilon^\mathrm{intra}(\omega)+\varepsilon^\mathrm{inter}(\omega).
\end{equation}
The imaginary part of the interband part can be calculated using the results from DFT calculations as\cite{Harl2008}
\begin{eqnarray}
\mathrm{Im}\left[\varepsilon^\mathrm{inter}_{\alpha\beta}(\omega)\right]&=&\frac{4\pi^2e^2}{V}\lim_{q\rightarrow 0}\frac{1}{q^2}\sum_{nm;\bm k}
2f_{n\bm k}\langle u_{m,\bm k+q\bm e_\alpha}|u_{n,\bm k}\rangle \langle u_{n,\bm k}|u_{m,\bm k+q\bm e_\beta}\rangle\cr
&&\qquad\qquad \times\left[\delta(E_{m,\bm k}-E_{n,\bm k}-\omega)-\delta(E_{m,\bm k}-E_{n,\bm k}+\omega)\frac{}{}\right],
\end{eqnarray}
where $\alpha$ and $\beta$ refer to Cartesian coordinates, $\bm e_{\alpha(\beta)}$ are unit vectors, $V$ is the volume of the unit cell, $|u_{n,\bm k}\rangle$ and $E_{n,\bm k}$ are the periodic part of the Bloch wave function and the corresponding eigenenergy for band $n$ and wave vector $\bm k$, and $f_{n\bm k}$ is the Fermi-Dirac distribution function. The real part of interband contribution can be obtained through the Kramers-Kronig relation. The intraband contribution is usually modeled by the Drude model:
\begin{equation}
\mathrm{Im}\left[\varepsilon^\mathrm{intra}_{\alpha\beta}(\omega)\right]=\frac{\gamma\omega^2_{p,\alpha\beta}}{\omega(\omega^2+\gamma^2)},
\end{equation}
\begin{equation}
\mathrm{Re}\left[\varepsilon^\mathrm{intra}_{\alpha\beta}(\omega)\right]=1-\frac{\omega^2_{p,\alpha\beta}}{\omega^2+\gamma^2}.
\end{equation}
Here $\gamma$ is a life-time broadening obtained either from a higher-order calculation or from experiments. In our calculation, we used the experimental determined electron life-time (of 0.6ps)\cite{Lee2013} for bulk Ca$_2$N to estimate $\gamma$ ($\sim1.1$meV). The same value was also used for the calculations of Sr$_2$N due to their similar electronic structures. We have checked the sensitivity of our results' dependence on $\gamma$ by repeating the calculations of $\varepsilon(\omega)$ with $\gamma$ value varying from $\gamma/10$ to $5\gamma$. The obtained results are of little difference. This is because the value of $\gamma$ is already quite small (reflecting the fact that these materials are good metals). The $\omega_{p,\alpha\beta}$ is the plasma frequency tensor which can be calculated using
\begin{equation}
\omega_{p,\alpha\beta}^2=-\frac{4\pi e^2}{V}\sum_{n;\bm k}2f'_{n\bm k}\left(\bm e_{\alpha}\cdot\frac{\partial E_{n,\bm k}}{\partial\bm k}\right)
\left(\bm e_{\beta}\cdot\frac{\partial E_{n,\bm k}}{\partial\bm k}\right).
\end{equation}
The dielectric functions and plasma frequencies were suitably renormalized to exclude the vacuum region from the unit cell in our calculations.

\subsection{Surface plasmon modes calculations.}
For thin film structures, the plasmon modes at two surfaces would couple and form two modes with different parity\cite{Dionne2005}. The equations governing their dispersions have been derived before\cite{Laref2013}:
\begin{equation}
L+:\qquad \varepsilon_{xx}\sqrt{\beta^2-\varepsilon_d k_0^2}-i\varepsilon_d\sqrt{\frac{\varepsilon_{xx}}{\varepsilon_{zz}}\beta^2-\varepsilon_{xx} k_0^2
}\tan\left(i\sqrt{\frac{\varepsilon_{xx}}{\varepsilon_{zz}}\beta^2-\varepsilon_{xx} k_0^2
}\frac{L}{2}\right)=0;
\end{equation}
\begin{equation}
L-:\qquad \varepsilon_{xx}\sqrt{\beta^2-\varepsilon_d k_0^2}+i\varepsilon_d\sqrt{\frac{\varepsilon_{xx}}{\varepsilon_{zz}}\beta^2-\varepsilon_{xx} k_0^2
}\cot\left(i\sqrt{\frac{\varepsilon_{xx}}{\varepsilon_{zz}}\beta^2-\varepsilon_{xx} k_0^2
}\frac{L}{2}\right)=0.
\end{equation}
The first equation above is for the antisymmetric ($L+$) mode and the second equation is for the symmetric ($L-$) mode. Here $L$ in the equations is the film thickness. Other quantities in these two equations are defined in the main text. In our calculation, we solve the two equations numerically using a two-dimensional unconstrained Nelder-Mead minimization algorithm\cite{Nelder1965} with a tolerance of $10^{-13}$nm$^{-1}$ in the complex wave vectors.


\bibliographystyle{apsrev4-1}

\begin{thebibliography}{100}

\bibitem{Dye2003} Dye, J.~L. Electrons as Anions. \emph{Science} \textbf{301}, 607-608 (2003).

\bibitem{Dye2009} Dye, J.~L. Electrides: early examples of quantum confinement. \emph{Acc.Chem.Res.} \textbf{42}, 1564-1572 (2009).

\bibitem{Ellaboudy1983} Ellaboudy, A., Dye, J.~L. \& Smith, P.~B. Cesium 18-Crown-6 Compounds. A Crystalline Ceside and a Crystalline Electride.
\emph{J. Am. Chem. Soc.} \textbf{105}, 6490-6491 (1983).

\bibitem{Huang1988} Huang, R.~H., Faber, M.~K., Moeggenborg, K.~J., Ward, D.~L. \& Dye, J.~L. Structure of K$^+$(cryptand[2.2.2J) electride and evidence for trapped electron pairs. \emph{Nature} \textbf{331}, 599-601 (1988).

\bibitem{Dye1990} Dye, J.~L. Electrides: Ionic Salts with Electrons as the Anions. \emph{Science} \textbf{247}, 663-668 (1990).

\bibitem{Matsuishi2003} Matsuishi, S. \emph{et al.} High-Density Electron Anions in a Nonaporous Single Crystal:[Ca$_{24}$Al$_{28}$O$_{64}$]$^{4+}$(4e$^-$). \emph{Science} \textbf{301}, 626 (2003).

\bibitem{Li2003} Li, Z.~Y., Yang, J.~L., Hou, J.~G. \& Zhu, Q.~S. Inorganic Electride: Theoretical Study on Structural and Electronic Properties. \emph{J. Am. Chem. Soc.} \textbf{125}, 6050-6051 (2003).

\bibitem{Toda2007} Toda, Y., Yanagi, H., Ikenaga, E., Kim, J.~J., Kobata, M., Ueda, S., Kamiya, T., Hirano, M., Kobayashi, K.\& Hosono, H. Work Function of a Room-Temperature, Stable Electride [Ca$_24$Al$_28$O$_64$]$^{4+}$(e$^-$)$_4$. \emph{Adv. Mater.} \textbf{19}, 3564-3569 (2007).

\bibitem{Kim2012} Kim, S.~W. \& Hosono, H. Synthesis and Properties of 12CaO$\cdot$7Al$_2$O$_3$ Electride: Review of Single Crystal and Thin Film Growth. \emph{Philos. Mag.} \textbf{92}, 2596 (2012).

\bibitem{Fang2000} Fang, C.~M., de Wijs, G.~A., de Groot, R.~A., Hintzen, H.~T. \& de With, G. Bulk and Surface Electronic Structure of the Layered Sub-Nitrides Ca$_2$N and Sr$_2$N, \emph{Chem. Mater.} \textbf{12}, 1847-1852 (2000).

\bibitem{Kitano2012} Kitano, M., Inoue, Y., Yamazaki, Y., Hayashi, F., Kanbara, S., Matsuishi, S., Yokoyama, T., Kim, S.~W., Hara, M. \& Hosono, H. Ammonia synthesis using a stable electride as an electron donor and reversible hydrogen store. \emph{Nat Chem} \textbf{4}, 934-940 (2012).

\bibitem{Picard2010} Pickard, C.~J. \& Needs, R.~J. Aluminium at Terapascal Pressures. \emph{Nat. Mater.} \textbf{9}, 624 (2010).

\bibitem{Gatti2010} Gatti, M., Tokatly, I.~V. \& Rubio, A. Sodium: A Charge-Transfer Insulator at High Pressures. \emph{Phys. Rev. Lett.} \textbf{104}, 216404 (2010).

\bibitem{Miao2014} Miao, M.-S. \& Hoffmann, R. High Pressure Electrides: A Predicative Chemical and Physical Theory. \emph{Acc. Chem. Res.} \textbf{47}, 1311 (2014).



\bibitem{Lee2013} Lee, K., Kim, S.~W., Toda, Y., Matsuishi, S \& Hosono, H. Dicalcium nitride as a two-dimensional electride with an anionic electron layer. \emph{Nature} \textbf{494}, 336-340 (2013).

\bibitem{Inoshita2014} Inoshita, T., Jeong, S., Hamada, N. \& Hosono, H. Exploration for Two-Dimensional Electrides via Database Screening and \textit{Ab Initio} Calculation. \emph{Phys. Rev. X.} \textbf{4}, 031023 (2014).

\bibitem{Tada2014} Tada, T., Takemoto, S., Matsuishi, S. \& Hosono, H. High-Throughput \textit{ab Initio} Screening for Two-Dimensional Electride Materials. \emph{Inorg. Chem.} \textbf{53}, 10347-10358(2014).

\bibitem{Walsh2013} Walsh, A., \& Scanlon, D. Electron excess in alkaline earth sub-nitrides: 2D electron gas or 3D electride? \emph{J. Mater. Chem. C} \textbf{1}, 3525-3528 (2013).

\bibitem{Zhao2014} Zhao, S.~T., Li, Z.~Y. \& Yang, J.~L. Obtaining two-dimensional electron gas in free space without resorting to electron doping: an electride based design. \emph{J. Am. Chem. Soc.} \textbf{136}, 13313-13318 (2014).



\bibitem{Smith2003} Smith, D.~R. \& Schurig, D. Electromagnetic Wave Propagation in Media with Indefinite Permittivity and Permeability Tensors. \emph{Phys. Rev. Lett.} \textbf{90}, 077405 (2003).

\bibitem{Gregory2000} Gregory, D.~H., Bowman, A., Baker, C.~F. \& Weston, D.~P. Dicalcium Nitride, Ca$_2$N-a 2D $"$Excess Electron$"$ Compound: Synthetic Routes and Crystal Chemistry. \emph{J. Mater. Chem.} \textbf{10}, 1635 (2000).

\bibitem{Baker2001} Baker, C.~F., Barker, M.~G., \& Blake, A.~J. Calcium nitride (Ca$_2$N), a redetermination. \emph{Acta Cryst.} E\textbf{57}, i6-i7 (2001).

\bibitem{Reckeweg2002} Reckeweg, O. \& DiSalvo, F.~J. Alkaline earth metal nitride compounds with the composition M$_2$NX (M=Ca, Sr, Ba; X=$\Box$, H, Cl or Br), \emph{Solid State Sciences} \textbf{4}, 575-584 (2002).


\bibitem{Grimme2006} Grimme, S.
Semiempirical GGA-type density functional constructed with a long-range dispersion correction. plasmonics.
\emph{J Comput Chem.} \textbf{27}, 1787 (2006).




\bibitem{Zabel2001} Zabel, H. Phonons in layered compounds. \emph{J. Phys.: Condens. Matter} \textbf{13}, 7679 (2001).

\bibitem{Liu2007a} Liu, F., Ming, P. \& Li, J. \textit{Ab initio} calculation of ideal strength and phonon instability of graphene under tension. \emph{Phys. Rev. B} \textbf{76}, 046120 (2007).

\bibitem{Sanchez2011} Molina-Sanchez, A. \& Wirtz, L. Phonons in single-layer and few-layer MoS$_2$ and WS$_2$. \emph{Phys. Rev. B} \textbf{84}, 155413 (2011).

\bibitem{Zhu2014} Zhu, L., Zhang, G. \& Li, B. Coexistence of size-dependent and size-independent thermal conductivities in phosphorene. \emph{Phys. Rev. B} \textbf{90}, 214302 (2014).



\bibitem{Zacharia2004} Zacharia, R., Ulbricht, H. \&  Hertel, T. Interlayer cohesive energy of graphite from thermal desorption of polyaromatic hydrocarbons. \emph{Phys. Rev. B} \textbf{69}, 155406 (2004).





\bibitem{Becke1990} Becke, A.~D. \& Edgecombe, K.~E. A simple measure of electron localization in atomic and molecular systems. \emph{The Journal of Chemical Physics} \textbf{92}, 5397-5403 (1990).

\bibitem{Silvi1994} Silvi, B. \& Savin, A. Classification of chemical bonds based on topological analysis of electron localization functions. \emph{Nature} \textbf{371}, 683-686 (1994).




\bibitem{Uijitewaal2004} Uijttewaal, M.~A., Wijs, G.~A.~de \& Groot, R.~A.~de. Low work function of the (1000) Ca$_2$N surface. \emph{J. Appl. Phys.} \textbf{96}, 1751-1753 (2004).




\bibitem{Harl2007} Harl, J., Kresse, G., Sun, L.~D., Hohage, M. \& Zeppenfeld, L. \textit{Ab initio} reflectance difference spectra of the bare and adsorbate covered Cu(110) surfaces. \emph{Phys. Rev. B} \textbf{76}, 035436 (2007).

\bibitem{Harl2008} Harl, J. The linear response function in density functional theory: Optical spectra and improved description of the electron correlation. \emph{ Ph.D Dissertation submitted to Universitat Wien} (2008).

\bibitem{He2010} He, Y. \& Zeng T. First-principle study and model of dielectric functions of silver nanoparticles. \emph{J. Phys. Chem. C} \textbf{114}(42), 18023-18030 (2010).

\bibitem{Glantschnig2010} Glantschnig, K. \& Ambrosch-Draxl, C. Relativistic effects on the linear optical properties of Au, Pt, Pb and W. \emph{New J. Phys.} \textbf{12}, 103048 (2010).

\bibitem{Yan2011} Yan, J., Jacobsen, K.~W. \& Thygesen, K.~Y. First principles study of surface plasmons on Ag(111) and H/Ag(111). \emph{Phys. Rev. B} \textbf{84}, 235430 (2011).

\bibitem{Laref2013} Laref, S., Cao, J.~R.,  Asaduzzaman, A., Runge, K., Deymier, P., Ziolkowski, R.~W., Miyawaki, M. \& Muralidharan, K. Size-dependent permittivity and intrinsic optical anisotropy of nanometric gold thin films: a density functional theory study. \emph{Opt. Express} \textbf{21}, 11827-11838 (2013).

\bibitem{Ming2014} Ming, W.~M., Blair, S. \& Liu, F. Quantum size effect on dielectric function of ultrathin metal film: a first-principles study of Al(111). \emph{J. Phys.: Condens. Matter} \textbf{26}, 505302 (2014).






\bibitem{Jacob2006} Jacob, Z., Alekseyev, L.~V. \& Narimanov, E. Optical hyperlens: Far-field imaging beyond the diffraction limit. \emph{Opt. Express} \textbf{14}, 8247-8256 (2006).

\bibitem{Liu2007} Liu, Z., Lee, H., Xiong, Y., Sun, C. \& Zhang, X. Far-field optical hyperlens magnifying sub-diffraction-limited objects. \emph{Science} \textbf{315}, 1686 (2007).





\bibitem{Maierbook} Maier, S.~A. \emph{Plasmonics: Fundamentals and Applications}. Springer, New York, 2007.

\bibitem{West2010} West, P.~R., Ishii, S., Naik, G.~V., Emani, N.~K., Shalaev, V.~M. \& Boltasseva, A. Searching for better plasmonics materials. \emph{Laser \& Photon. Rev.} \textbf{4}, 795-808 (2010).

\bibitem{Dionne2005} Dionne, J.~A., Sweatlock, L.~A., Atwater, H.~A. \& Polman, A. Planar metal plasmon waveguides: frequency-dependent dispersion, propagation, localization, and loss beyond the free electron model. \emph{Phys. Rev. B} \textbf{72}, 075405 (2005).




\bibitem{Kresse1993} Kresse, G. \& Hafner, J. \textit{Ab initio} molecular dynamics for open-shell transition metals. \emph{Phys. Rev. B} \textbf{48}, 13115-13118 (1993).
\bibitem{Kresse1996} Kresse, G. \& Furthm\"uller, J. Efficient iterative schemes for \textit{ab initio} total-energy calculations using a plane-wave basis set. \emph{Phys. Rev. B} \textbf{54}, 11169-11186 (1996).

\bibitem{Perdew1996} Perdew, J., Burke, K. \& Ernzerhof, M. Generalized Gradient Approximation Made Simple. \emph{Phys. Rev. Lett.} \textbf{77}, 3865-3868 (1996).

\bibitem{Bloechl1994} Bl\"ochl, P. Projector augmented-wave method. \emph{Phys. Rev. B} \textbf{50}, 17953-17979 (1994).


\bibitem{Monkhorst1976} Monkhorst, H. \& Pack, J. Special points for Brillouin-zone integrations. \emph{Phys. Rev. B} \textbf{13}, 5188-5192 (1976).



\bibitem{Togo2008} Togo, A., Oba, F. \& Tanaka, I. First-principles calculations of the ferroelastic transition between rutile-type and CaCl$_2$-type SiO$_2$ at high pressures. \emph{Phys. Rev. B} \textbf{78}, 134106 (2008).

\bibitem{Gonze1997} Gonze, X.\& Lee, C.~Y. Dynamical matrices, Born effective charges, dielectric permittivity tensors, and interatomic force constants from density-functional perturbation theory. \emph{Phys. Rev. B} \textbf{55}, 10355-10368 (1997).

\bibitem{Methfessel989} Methfessel, M. \& Paxton, A. High-precision sampling for Brillouin-zone integration in metals. \emph{Phys. Rev. B} \textbf{40}, 3616-3621 (1989).

\bibitem{Nelder1965} Nelder, J.~A. \& Mead, R. A Simplex Method for Function Minimization. \emph{The Computer Journal} \textbf{7}, 308-313 (1965).


\end{thebibliography}

\begin{addendum}
\item [Acknowledgements]
The authors thank D.L. Deng for helpful discussions. This work was supported by the MOST Project
of China (Nos. 2014CB920903, 2011CBA00100), the NSF of China (Nos. 11174337, 11225418) , the SRFDPHE of China (No. 20121101110046) and the SUTD-SRG-EPD2013062.

\item [Author Contributions]
Y.Y. and S.A.Y. conceived the idea and supervised the work. S.G. performed the calculation and the data analysis.
S.G., L.Z., S.A.Y., and Y.Y. contributed to the interpretation of the results and
wrote the manuscript. All authors contributed in the discussion and reviewed the manuscript.

\item [Competing Interests]
The authors declare no competing financial interests.

\item [Correspondence]
Correspondence should be addressed to Shengyuan A. Yang or Yugui Yao.

\end{addendum}

\clearpage

\newpage
\begin{table}
 \caption{\label{lattice} Lattice parameters of Ca$_2$N and Sr$_2$N thin films with thicknesses from 1-ML to 5-ML. Here $a$ is the in-plane lattice constant, $t$ is the thickness of the film (vertical distance between the top and bottom atomic layers), L(1,2,3) refer to the thicknesses of the [Ca$_2$N]$^+$ layers, L1 is for the outermost layer(s), L2 is for the next outermost layer(s) and so on (see Fig.1), and G$ij$ refers to the thickness of the interlayer gap region between Layer $i$ and Layer $j$.}
  \begin{tabular}{cccccccc}
   \multicolumn{8}{c}{} \\
   \hline\hline
   \multicolumn{1}{c}{\textrm{No. of Layers}} &
   \multicolumn{1}{c}{\textrm{a (\AA)}} &
   \multicolumn{1}{c}{\textrm{t (\AA)}} &
   \multicolumn{1}{c}{\textrm{L1 (\AA)}} &
   \multicolumn{1}{c}{\textrm{L2 (\AA)}} &
   \multicolumn{1}{c}{\textrm{L3 (\AA)}} &
   \multicolumn{1}{c}{\textrm{G12 (\AA)}} &
   \multicolumn{1}{c}{\textrm{G23 (\AA)}} \\
   \hline\hline
   \multicolumn{1}{c}{\textrm{Ca$_2$N}} &
   \multicolumn{7}{c}{\textrm{}} \\
  1 & 3.562 & 2.516 & 2.516 & & & & \\
  2 & 3.562 & 8.608 & 2.509 & & & 3.598 & \\
  3 & 3.552 & 14.957 & 2.514 & 2.504 & & 3.712 & \\
  4 & 3.550 & 20.099 & 2.516 & 2.507 & & 3.721 & 3.793\\
  5 & 3.555 & 26.419 & 2.513 & 2.505 & 2.508 & 3.703 & 3.823 \\
   \hline
   \multicolumn{1}{c}{\textrm{Sr$_2$N}} &
   \multicolumn{7}{c}{\textrm{}} \\
  1 & 3.778 & 2.761 & 2.761 & & & & \\
  2 & 3.775 & 9.427 & 2.757 & & & 3.913 & \\
  3 & 3.770 & 16.260 & 2.760 & 2.748 & & 3.983 & \\
  4 & 3.765 & 21.821 & 2.764 & 2.754 & & 4.016 & 4.068\\
  5 & 3.768 & 28.562 & 2.758 & 2.751 & 2.753 & 3.989 & 4.061 \\
   \hline
   \end{tabular}
\end{table}

\newpage

\newpage

\newpage

\newpage

\newpage

\newpage

\newpage

\newpage

\newpage

\newpage

\newpage

\newpage

\newpage

\newpage

\newpage

\newpage

\newpage

\end{document}